\documentclass[a4paper,11pt]{article}
\usepackage{jheppub}
\usepackage{orcidlink}
\usepackage{caption}
\usepackage{subcaption}
\usepackage{setspace}
\usepackage{hyperref}
\title{\boldmath Cylindrical black hole solutions in $f(\mathcal{R})$ and $f(\mathcal{R},\mathcal{A},A^{\mu\nu}A_{\mu\nu})$ modified gravity}

\author[a]{Faizuddin Ahmed\orcidlink{0000-0003-2196-9622}}
\author[b]{and Abdelmalek Bouzenada\orcidlink{0000-0002-3363-980X}}
\affiliation[a]{Department of Physics, University of Science \& Technology Meghalaya, Ri-Bhoi, Meghalaya, 793101, India}
\affiliation[b]{Laboratory of Theoretical and Applied Physics, Echahid Cheikh Larbi Tebessi University 12001, Algeria}

\emailAdd{faizuddinahmed15@gmail.com}
\emailAdd{abdelmalekbouzenada@gmail.com}

\abstract{
In this paper, we explore a cylindrical black hole (BH) space-time introduced by Lemos (Phys. Lett. {\bf B 353}, 46 (1995)), in the context of modified gravity theories. Specifically, we focus on $f(\mathcal{R})$-gravity framework, where we choose two form functions, $f(\mathcal{R})=(\mathcal{R}+\alpha_1\,\mathcal{R}^2+\alpha_2\,\mathcal{R}^3+\alpha_3\,\mathcal{R}^4+\alpha_4\,\mathcal{R}^5)$ and $f(\mathcal{R})=\mathcal{R}+\alpha_k\,\mathcal{R}^{k+1},\quad (k=1,2,...n)$. We solve the modified field equations incorporating zero energy-momentum tensor, $\mathcal{T}^{\mu\nu}=0$ and obtain the result. Moreover, we study another well-known modified gravity theory called Ricci-Inverse ($\mathcal{RI}$) gravity and investigate this Lemos black hole (LBH) space-time. To achieve this, we consider different classes of models defined as follows: (i) Class-\textbf{I} model: $f(\mathcal{R}, \mathcal{A})=(\mathcal{R}+\beta\,\mathcal{A})$, (ii) Class-\textbf{II} model: $f(\mathcal{R}, A^{\mu\nu}\,A_{\mu\nu})=(\mathcal{R}+\gamma\,A^{\mu\nu}\,A_{\mu\nu})$, and (iii) Class-\textbf{III} model: $f(\mathcal{R}, \mathcal{A}, A^{\mu\nu}\,A_{\mu\nu})=(\mathcal{R}+\alpha_1\,\mathcal{R}^2+\alpha_2\,\mathcal{R}^3+\beta_1\,\mathcal{A}+\beta_2\,\mathcal{A}^2+\gamma\,A^{\mu\nu}\,A_{\mu\nu})$, where $\mathcal{A}=g_{\mu\nu}\,A^{\mu\nu}$ is the anti-curvature scalar, $A^{\mu\nu}$ is the anti-curvature tensor, the reciprocal of the Ricci tensor $R_{\mu\nu}$. We solve the modified field equations under the same aforementioned scenario of energy-momentum tensor, and obtain the result. Subsequently, we study the geodesic motions of test particles around this LBH within the Ricci-Inverse and $f(\mathcal{R})$-gravity theories and analyze the outcomes. We demonstrate that different coupling constants chosen in these modified gravity theories influences the usual cosmological constant $\Lambda$, and thus, shifted the result in comparison to the general relativity case. Moreover, we demonstrate that geodesics motions are influenced by these modified gravity and changes the results in comparison to general relativity. 
}


\begin{document}
\maketitle
\flushbottom

\section{Introduction}

General Relativity (GR) \cite{k1} is one of the cornerstones of modern physics, providing insights into the universe from fundamentally different scales and perspectives. Formulated by Albert Einstein in the early 20th century, GR revolutionized our understanding of gravity by positing that massive objects warp the fabric of spacetime, resulting in what we perceive as gravitational attraction. This theory has yielded profound insights into large-scale cosmic structures and events, including the dynamics of black holes, the formation of galaxies, and the overall expansion of the universe \cite{k3}. GR is founded on two main principles: the principle of general covariance and the principle of equivalence. The principle of general covariance asserts that the laws of physics must be the same for all observers, motivating the use of tensors to describe gravity. Tensors are invariant under coordinate transformations, meaning that while the components of a vector may change when the coordinate system is altered, the vector itself continues to point in the same direction. The equivalence principle states that inertial and gravitational masses are identical, implying that all objects fall with the same acceleration in a gravitational field, regardless of their mass. The strongest form of this principle, known as the Einstein Equivalence Principle (EEP), posits that all physical laws reduce to those of special relativity in any inertial (freely falling) frame, provided that the region of space considered is small compared to the length scale over which the gravitational field varies.

Since the initial discovery of classical black hole (BH) solutions \cite{n1,n2,n3,n4}, a significant body of research has emerged to investigate the complex dynamics of photons and other particles in the extreme gravitational environments surrounding these cosmic objects \cite{n6}. One of the most intriguing aspects of this research is the behavior of light in the presence of strong gravitational fields, which leads to the bending of light paths, a phenomenon known as gravitational lensing. The curvature of spacetime near black holes is so intense that it can dramatically alter the trajectory of photons. This bending effect is especially pronounced in regions such as the photon sphere, the critical boundary where photons can be trapped in unstable circular orbits around the black hole \cite{n7,n8}. These areas are not only theoretically captivating but also serve as prime observational targets for testing high-energy physics in the most extreme gravitational environments known to us \cite{n9}.

The existence of BHs is arguably one of the most astonishing predictions of GR and dates back to the first BH solution to the EFEs: the Schwarzschild solution, which introduced what is now known as the Schwarzschild metric \cite{n1}. This metric is given in Schwarzschild coordinates by:
\begin{equation}
    ds^2=-\left(1-\frac{2\,G\,M/c^2}{r} \right)\,dt^2+\left(1-\frac{2\,G\,M/c^2}{r} \right)^{-1}\,dr^2+r^2\,(d\theta^2+\sin^2 \theta\,d\phi^2),\label{Sch}
\end{equation}
where $M$ is a positive constant that represents the mass of the BH, and $r,\theta,\phi$ are standard spherical coordinates. The metric is time-independent and solves the vacuum Einstein Field Equations (EFE). It represents the most general spherically symmetric solution to the EFE, which is necessarily static and asymptotically flat, as stated in Birkhoff’s theorem \cite{GDB}. Notably, the Schwarzschild solution can only be considered the most general solution if it encompasses Minkowski space, which is obtained by setting $M$ to zero. The Schwarzschild metric describes the curvature of spacetime around a spherically symmetric central mass and features a physical singularity at the origin, where the Ricci scalar approaches infinity. It is important to note that the singularity at $r=2\,M$ in Eqn. (\ref{Sch}) arises from the choice of coordinates and is therefore not physical. The existence of an object such as a black hole, with its singularity and corresponding event horizon, is indeed a bold proposition that has gained widespread acceptance. There is substantial experimental evidence supporting the existence of black holes. For example, the orbits of stars in the Galactic Center have been utilized to map the gravitational potential in that region, which corresponds to that of a supermassive black hole \cite{KK1,KK2,KK3}. Additionally, gravitational wave observations and the image of the black hole in M87* provide further compelling evidence for their existence.

The lensing effect, as described by general relativity, is intricately dependent on the relative positioning of the light source, the black hole, and the observer. Specifically, the alignment concerning the optical axis is crucial in determining the configuration of the gravitational lensing system. When this alignment is nearly perfect, the lensing effect can produce phenomena such as Einstein rings or multiple images of the same distant object, resulting from the deflection of light around the black hole. For observers, studying the bending of light near the photon sphere offers a remarkable opportunity to investigate the interaction between light and gravity in unprecedented detail \cite{n10}. Moreover, numerous studies have investigated lensing effects at large deflection angles, particularly in scenarios involving minimal coupling between photons and the gravitational field. These investigations demonstrate that, under such conditions, photons can approach the photon sphere with remarkable proximity, often circling the black hole multiple times before either escaping or being captured by the event horizon. This repeated orbiting of photons, known as strong lensing, results in highly magnified and distorted images of the light source. The intricate behavior of photons in this regime offers critical insights into both the geometry of spacetime and the nature of black hole singularities \cite{n11,n12}.

In addition to its fundamental implications for gravitational lensing, the strong deflection limit near the photon sphere provides a valuable opportunity to probe deeper aspects of black hole physics, including the nature of event horizons and potential quantum gravitational effects \cite{n13}. The extreme gravitational environment surrounding black holes may unveil phenomena that extend beyond the predictions of classical general relativity. Furthermore, black holes act as ideal laboratories for testing the interplay between quantum mechanics and gravity. For example, deviations from classical predictions in the behavior of light near the photon sphere could signal new physics, such as quantum corrections to the gravitational field or the influence of higher-dimensional spacetime geometries \cite{n14}. The possibility of photons orbiting the black hole multiple times before reaching an observer introduces a rich array of observational signatures, including the formation of photon rings—concentric circles of light that reveal the intricate geometry of the black hole’s event horizon \cite{n15}. Such observations have the potential to illuminate fundamental questions about the nature of spacetime, the behavior of gravity at small scales, and even the possible existence of new particles or fields that interact with photons in these high-energy environments \cite{n16}.

Black holes are enigmatic cosmic phenomena characterized by their immense gravitational pull and intricate structures, offering critical insights into the nature of gravity, particularly within the framework of General Relativity. According to this theory, black holes are defined as regions in spacetime that are bounded by an event horizon. Notably, Stephen Hawking demonstrated that the size of this event horizon cannot decrease over time \cite{Hawking1972}. In a groundbreaking contribution, Jacob Bekenstein introduced the concept of black hole entropy \cite{Bekenstein1972, Bekenstein1973, Bekenstein1974}, establishing a connection between the thermodynamic properties of black holes and the principles of quantum mechanics. Further investigations revealed a significant relationship between a black hole's surface gravity and its temperature ($T_{H}$), indicating that the surface gravity influences the thermal characteristics of the black hole \cite{Bardeen1973}. Building on this foundation, Hawking proposed that black holes can emit thermal radiation, a phenomenon now known as Hawking radiation. This radiation arises from quantum fluctuations near the event horizon, where pairs of virtual particles are generated. In this process, one particle escapes as radiation, while the other falls into the black hole. This results in a detectable outflow of positive energy, fundamentally altering our understanding of black hole thermodynamics \cite{Hawking1974}. The implications of these findings are profound, as they bridge quantum mechanics and gravity, reshaping our perception of black holes not just as voids of gravitational pull, but as dynamic entities with thermodynamic properties. This ongoing dialogue between the theories of thermodynamics, quantum mechanics, and gravity continues to challenge and enrich our comprehension of the universe.

Theoretical investigations are actively exploring methods to detect black holes in the universe \cite{Abbott2016, Akiyama2019a, Akiyama2019b, Abbott2020}. This research has gained substantial validation through observational evidence, such as the detection of gravitational waves and the groundbreaking optical imaging of a black hole's shadow. As a result, black holes are increasingly conceptualized as thermodynamic systems, where their temperature is linked to surface gravity and their entropy is proportional to the area of the event horizon. This perspective has led to the identification of the Hawking-Bekenstein phase transition, which describes a shift from a state dominated by pure thermal radiation to the formation of a stable, massive Schwarzschild black hole in anti-de Sitter (AdS) space \cite{Hawking1983, Witten1998}.  Furthermore, research indicates a fundamental phase transition in charged AdS black holes, reminiscent of the liquid-gas transition observed in classical thermodynamic systems. This analogy highlights intriguing parallels with well-established thermodynamic phenomena \cite{Kubiznak2012}. Many studies are dedicated to elucidating the phase behavior in various types of black hole theories. This line of inquiry is crucial for advancing our understanding of black hole thermodynamics in these extreme gravitational environments \cite{PV1, PV2, PV3, PV4}. The ongoing exploration of these topics not only deepens our knowledge of black holes but also enriches the broader discourse on the interplay between thermodynamics and gravity.

Modified gravity models present diverse strategies to explain dark energy, including notable frameworks like $f(R)$ gravity, $f(R,T)$ gravity, $f(\mathcal{G})$ gravity, $f(R,\mathcal{G})$ gravity, and $f(R,T_{\mu\nu}T^{\mu\nu})$ gravity, among others. Among these, $f(R)$ gravity has gained significant attention, particularly for its implications in high-curvature regimes, achieved by substituting the Ricci scalar $R$ with a custom function of $R$ in the Einstein-Hilbert action \cite{a1,a2,a3,a4,a5, HHH1, HHH2, HHH3, HHH4}. While it shares some solutions with general relativity (GR), $f(R)$ gravity also offers unique solutions with distinct physical properties, making the exploration of black hole (BH) solutions in this context particularly crucial. However, the presence of higher-order derivative terms complicates solving the motion equations. Various approaches have been proposed to address these challenges (e.g., \cite{a6}), including solutions for black holes in $f(R)$ gravity interacting with nonlinear electromagnetic fields \cite{a7,a8,a9,a10}. Some studies of $f(R)$-gravity theory in various context, see references \cite{GM1,GM2,GM3,GM4,GM5,GM6,GM7} and other studies in Refs. \cite{GM8,GM9}. These solutions not only provide valuable opportunities to test the theoretical predictions of $f(R)$ gravity against observational data but also enhance our understanding of gravity and its effects on astrophysical and cosmological scales.

In $f(R)$ gravity, there are three propagating degrees of freedom, which include an additional scalar polarization mode alongside the two tensor polarizations found in General Relativity \cite{aa}. This scalar polarization comprises a combination of longitudinal and breathing modes. When the scalar mass of $f(R)$ is zero, the longitudinal mode disappears, leaving only the breathing mode \cite{aa1}. Generally, a positive scalar mass is necessary to ensure the stability of cosmological perturbations \cite{aa2,aa3}, leading to the conclusion that the massless case should be dismissed. However, stable perturbations can still arise in the massless scenario if certain additional constraints are considered \cite{aa4}. To comprehensively understand the polarization behavior of $f(R)$ gravity, it is crucial to establish these detailed constraints and reevaluate the potential for stable massless scalar polarization.

Scalar-tensor theories \cite{bb1} and Brans-Dicke theory \cite{bb2}, which are largely equivalent to $f(R)$ gravity \cite{bb3,bb4}, do allow for massless scalar polarization; however, this feature is seldom observed in $f(R)$ gravity. This discrepancy highlights the need for further investigation into massless scalar polarization within $f(R)$ gravity. Although some studies have tackled this issue \cite{bb5}, several challenges remain. For instance, \cite{bb5} focuses on a specific power-law $f(R)$ model that lacks the essential characteristics of viable $f(R)$ models. It sets the scalar mass to zero directly but fails to address the necessary constraints for stable perturbations, which may leave the de Sitter point in the effective potential as a possible inflection point or local maximum, rather than the required local minimum. This gap calls for a more comprehensive exploration of the conditions under which stable massless scalar polarization can emerge in $f(R)$ gravity.

Another gravitational theory, known as Ricci-Inverse gravity \cite{ref1,ref2,ref3}, has recently gained significant attention in the context of modified gravity theories. This approach modifies the Einstein-Hilbert action by incorporating a term that is equal to the inverse of the Ricci tensor, $R_{\mu\nu}$, a quantity that encapsulates the curvature of spacetime. In the Einstein-Hilbert action, the Ricci scalar is replaced by a function $f(\mathcal{R}, \mathcal{A}, A^{\mu\nu}\,A_{\mu\nu})$, where $A^{\mu\nu}=R^{-1}_{\mu\nu}$ is the symmetric second-rank anti-curvature tensor, and $\mathcal{A}=g_{\mu\nu}\,A^{\mu\nu}$ is its scalar. Recently, several authors have investigated Ricci-Inverse gravity in various contexts, including models of causality violations, wormholes, and stellar structure, as reported in Refs. \cite{ref4,ref5,ref8,ref9,epjc,plb,meer,AM1,AM2,AM3,AM4, MFS2,cjph,ijtp,aop,cjphy,EPJP,NA,NPB,EPJC,EPJC2,EPJC4,JCAP}.

The primary objective of this work is to explore the Lemos cylindrical black hole (BH) spacetime formulated in general relativity theory, as originally proposed by Lemos \cite{JPSL}, within the context of modified gravity theories. We aim to establish that this Lemos black hole (LBH) serves as a valid solution in the context of the $f(\mathcal{R})$-gravity framework and analyze how the higher-order curvature terms influence the dynamics of the system. For this purpose, we choose two different functional forms given by:
$(\mathcal{R}+\alpha_1\,\mathcal{R}^2+\alpha_2\,\mathcal{R}^3+\alpha_3\,\mathcal{R}^4+\alpha_4\,\mathcal{R}^5)$ and $\mathcal{R}+\alpha_k\,\mathcal{R}^{k+1}\quad (k=1,2,...n)$.  Moreover, we consider another modified gravity called the Ricci-Inverse gravity of Class-\textbf{I} to Class-{\bf III} models defined in \cite{ref2,ref3}, such as (i) $f(\mathcal{R}, \mathcal{A})=(\mathcal{R}+\beta\,\mathcal{A})$, (ii) $f(\mathcal{R}, A^{\mu\nu}\,A_{\mu\nu})=(\mathcal{R}+\gamma\,A^{\mu\nu}\,A_{\mu\nu})$, and (iii) $f(\mathcal{R}, \mathcal{A}, A^{\mu\nu}\,A_{\mu\nu})=(\mathcal{R}+\alpha_1\,\mathcal{R}^2+\alpha_2\,\mathcal{R}^3+\beta_1\,\mathcal{A}+\beta_2\,\mathcal{A}^2+\gamma\,A^{\mu\nu}\,A_{\mu\nu})$. We solve the modified field equations in both gravity theories, incorporating vacuum as the matter-energy content, and discuss the results. Subsequently, we study the geodesic motions of test particles around this LBH within the framework of these modified gravity theories and analyze the outcomes. In fact, it is demonstrated that various coupling constants in modified gravity theories influence the geodesic paths of the system, leading to results that differ from those of general relativity.

The paper is organized as follows: In Section \ref{sec:2}, we investigate a cylindrical black hole spacetime within the framework of general relativity. This section provides a detailed examination of this specific BH spacetime in the context of the $f(\mathcal{R})$-gravity framework. In Section \ref{sec:3}, we explore this cylindrical black hole spacetime in the context of another modified gravity theory known as Ricci-Inverse gravity theory, covering models from Class \textbf{I} to Class \textbf{III}. Section \ref{sec:4} is dedicated to discussing the geodesic motions of test particles around this BH metric in both modified gravity theories, followed by an analysis of the results. Finally, Section \ref{sec:5} presents our conclusions. We choose the system of units where $\hbar=c=G=1$.

\section{Lemos BH solution within $f(\mathcal{R})$-gravity framework}\label{sec:2}

In this section, we consider a cylindrical black hole solution formulated within the general relativity, in the context of modified gravity theory. Therefore, we begin this section by introducing this BH solution in cylindrical coordinates ($t, r, \varphi, z$) given by \cite{JPSL}
\begin{equation}
ds^2=-f(r)\,dt^2+\frac{dr^2}{f(r)}+r^2d\varphi+\alpha^2\,r^2\,dz^2,\quad f(r)=\left(\alpha^2\,r^2-\frac{4\,M}{\alpha\,r} \right),\label{a1}
\end{equation}
where the metric tensor $g_{\mu\nu}$ and its contravairant form are given by $(x^0=t, x^1=r, x^2=\varphi, x^3=z)$
\begin{eqnarray}
g_{\mu\nu}=\left(\begin{array}{cccc}
     -f(r) & 0 & 0 & 0 \\
     0 & \frac{1}{f(r)} & 0 & 0\\
     0 & 0 & r^2 & 0\\
     0 & 0 & 0 & \alpha^2\,r^2 
\end{array}\right),\quad 
g^{\mu\nu}=\left(\begin{array}{cccc}
     -\frac{1}{f(r)} & 0 & 0 & 0\\
     0 & f(r) & 0 & 0\\
     0 & 0 & \frac{1}{r^2} & 0\\
     0 & 0 & 0 & \frac{1}{\alpha^2\,r^2}
\end{array}\right).\label{a2}
\end{eqnarray}
Here $M$ denotes mass of BH, $\alpha>0$ is a positive constant, and the coordinates are in the ranges $- \infty <t < \infty$, $r \geq 0$, $\varphi \in[0, 2\,\pi)$, and $- \infty < z < \infty$ called the temporal, radial, angular and axial coordinates, respectively.

For this metric (\ref{a1}), we calculate the Ricci tensor $R_{\mu\nu}$ and is given by
\begin{eqnarray}
R_{\mu\nu}=\left(\begin{array}{cccc}
     3\,\alpha^2\,\left(\alpha^2\,r^2-\frac{4\,M}{\alpha\,r} \right)& 0 & 0 &0 \\
     0& -\frac{3\,\alpha^2}{\left(\alpha^2\,r^2-\frac{M}{\alpha\,r} \right)} & 0 & 0\\
     0 & 0 & -3\,\alpha^2\,r^2 & 0\\
     0 & 0 & 0 & -3\,\alpha^4\,r^2
\end{array}\right).\label{a3}
\end{eqnarray}
The various scalar quantities, such as the Ricci scalar for this metric (\ref{a1}) is given by
\begin{equation}
    \mathcal{R}=g_{\mu\nu}\,R^{\mu\nu}=-12\,\alpha^2.\label{a4}
\end{equation}
And the Kretschmann scalar is given by
\begin{equation}
\mathcal{K}=R^{\mu\nu\lambda\sigma}\,R_{\mu\nu\lambda\sigma}=24\,\left(\alpha^4+\frac{8\,\mu^2}{\alpha^2\,r^6} \right).\label{a6}
\end{equation}
At infinite radial distance, $r \to \infty$, this Kretschmann scalar approaches a finite value, $\mathcal{K} \to 24\,\alpha^4$. In contrast, at the origin $r=0$, the scalar diverges, indicating the presence of a curvature singularity.

Now, the Einstein vacuum field equations with a cosmological constant $\Lambda$ are given by ($c=1=G$)
\begin{equation}
    R_{\mu\nu}=\Lambda\,g_{\mu\nu},\quad\quad \mathcal{R}=4\,\Lambda\,.\label{a7}
\end{equation}
Therefore, by substituting the metric tensor $g_{\mu\nu}$ from Eq. (\ref{a2}) and the Ricci tensor $R_{\mu\nu}$ from Eq. (\ref{a4}) into the field equations (\ref{a7}), one can find
\begin{equation}
    \Lambda=-3\,\alpha^2<0.\label{a8}
\end{equation}

Therefore, the cylindrical space-time described by the line-element (\ref{a1}) fulfills the vacuum field equations with a negative cosmological constant in the context of general relativity representing an anti-de Sitter space background.

Now, we examine black hole solution (\ref{a1}) within the framework of $f(\mathcal{R})$-gravity theory. In this approach, the Ricci scalar $\mathcal{R}$ is replaced by a function of itself, $f(\mathcal{R})$, in the Einstein-Hilbert action. Consequently, the action that characterizes $f(\mathcal{R})$-gravity is expressed as follows:
\begin{equation}\label{CC1}
    S = \int \mathrm{d}x^4 \sqrt{-g}\,[f(\mathcal{R})-2\,\Lambda ]+ S_M,\quad \mathcal{R}=g_{\mu\nu}\,R^{\mu\nu}.
\end{equation}

By varying the action (\ref{CC1}) with respect to the metric tensor $g_{\mu\nu}$, the modified field equations in $f(\mathcal{R})$-gravity are given by  
\begin{equation}\label{CC2}
    -\frac{1}{2}\,f(\mathcal{R})\,g^{\mu \nu} + f_{\mathcal{R}} \, R^{\mu \nu}+g^{\mu \nu} \nabla ^2 \, f_{\mathcal{R}}-\nabla ^{\mu} \nabla^{\nu}\, f_{\mathcal{R}}+ \Lambda_m\, g^{\mu \nu} =\mathcal{T}^{\mu \nu}.
\end{equation}

Let us consider the function $f$ in $f(\mathcal{R})$-gravity to be the following form:
\begin{equation} \label{CC3}
    f(\mathcal{R}) =\mathcal{R}+\alpha_1\,\mathcal{R}^2+\alpha_2\,\mathcal{R}^3+\alpha_3\,\mathcal{R}^4+\alpha_4\,\mathcal{R}^5, 
\end{equation}
where $\alpha_i\quad (i=1,..,4)$ are the coupling constants. As is well-know, the metric tensor $g_{\mu\nu}$ is dimensionless, while the Ricci tensor ($R_{\mu}$), the Ricci scalar ($R$) and the usual cosmological constant ($\Lambda$) all have dimensions of the inverse square of length, {\it i. e.}, $[L^{-2}]$. Therefore, the dimensions of the various coupling constants are as follows: $\alpha_1$ has dimension $[L^2]$, $\alpha_2$ has dimension $[L^4]$, $\alpha_3$ has dimension $[L^6]$, and $\alpha_4$ has dimension $[L^8]$, in the natural units where $G=1=c$. 

Using this function (\ref{CC3}), one can find the following
\begin{eqnarray}\label{CC4}
\partial f/\partial \mathcal{R}=f_{\mathcal{R}}=1+2\,\alpha_1\,\mathcal{R}+3\,\alpha_2\,\mathcal{R}^2+4\,\alpha_3\,\mathcal{R}^3+5\,\alpha_4\,\mathcal{R}^4.
\end{eqnarray}

Substituting the Ricci tensor $R_{\mu\nu}$ and the Ricci scalar $\mathcal{R}$ from Eq.(\ref{a7}), and the anti-curvature tensor $A^{\mu\nu}$ from Eq. (\ref{b2}) into the modified fields equations (\ref{CC2}), we obtain 

\begin{eqnarray}\label{CC5}
    &&\mathcal{T}^{\mu\nu}=\tilde{\tilde{\Lambda}}\,\begin{pmatrix}
        -\left(\alpha^2\,r^2-\frac{4\,M}{\alpha\,r} \right)^{-1} & 0 & 0 & 0\\
        0 & \left(\alpha^2\,r^2-\frac{4\,M}{\alpha\,r} \right) & 0 & 0\\
        0 & 0 & \frac{1}{r^2} & 0\\
        0 & 0 & 0 & \frac{1}{\alpha^2\,r^2}
    \end{pmatrix}=\tilde{\tilde{\Lambda}}\,g^{\mu\nu},\nonumber\\
    &&\tilde{\tilde{\Lambda}}=\left(\Lambda_m+3\,\alpha^2-432\,\alpha^6\,\alpha_2+10368\,\alpha^8\,\alpha_3-186624\,\alpha^{10}\,\alpha_4\right).
\end{eqnarray}

To solve the modified field equations (\ref{CC5}) in a vacuum where the matter-energy content is represented by an energy-momentum tensor of $T^{\mu\nu}=0$, we can simplify Eq. (\ref{CC5})
\begin{equation}
    \Lambda^{f(\mathcal{R})}_m=-3\,\alpha^2+432\,\alpha^6\,\alpha_2-10368\,\alpha^8\,\alpha_3+186624\,\alpha^{10}\,\alpha_4.\label{CC6}
\end{equation}

In terms of usual cosmological constant $\Lambda$, we can write the effective cosmological constant
\begin{equation}
    \Lambda^{f(\mathcal{R})}_m=\Lambda-16\,\Lambda^3\,\alpha_2-128\,\Lambda^4\,\alpha_3-768\,\alpha_4\,\Lambda^5.\label{CC7}
\end{equation}

The effective cosmological constant $\Lambda_m$ is negative provided the following condition must obey:
\begin{equation}\label{CC8}
    16\,\Lambda^3\,\alpha_2+768\,\alpha_4\,\Lambda^5 <\Lambda-128\,\Lambda^4\,\alpha_3.
\end{equation}

\begin{figure}[ht!]
    \centering
    \includegraphics[width=0.75\linewidth]{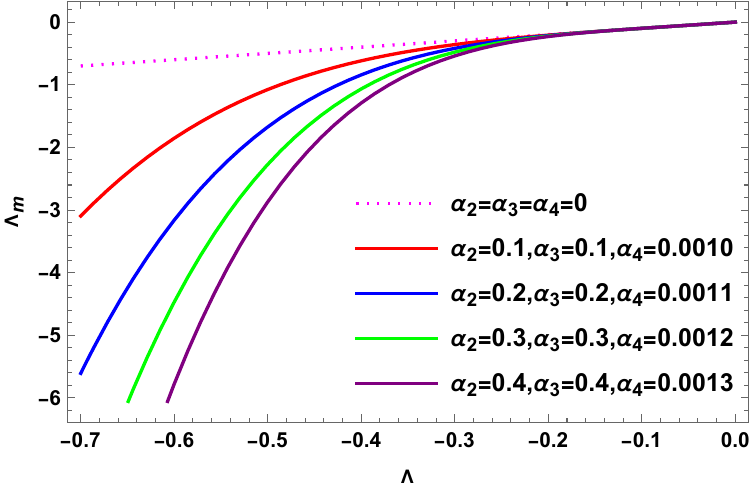}
    \caption{The effective cosmological constant $\Lambda_m$ as a function of the usual cosmological constant $\Lambda$.}
    \label{fig:2}
\end{figure}

In Figure \ref{fig:2}, we illustrate the effective cosmological constant $\Lambda_m$ as a function of the usual cosmological constant $\Lambda$. The figure showcases the variations of $\Lambda_m$ for different values of the coupling constants $\alpha_2,\alpha_3,\alpha_4$. The dotted red line represents the cases in general relativity, while the solid lines correspond to various values of the coupling constants. This visualization underscores the impact of these coupling constants on the relationship between the effective and traditional cosmological constants, offering valuable insights into the dynamics of the model.

The BH solution (\ref{a1}) in terms of effective cosmological constant $\Lambda^{f(\mathcal{R})}_m$ in $f(\mathcal{R})$-gravity framework is therefore described by the following line-element
\begin{eqnarray}
ds^2=\left(\frac{\Lambda^{f(\mathcal{R})}_m}{3}\,r^2+\frac{4\,M}{\sqrt{-\frac{\Lambda^{f(\mathcal{R})}_m}{3}}\,r}\right)\,dt^2-\frac{dr^2}{\left(\frac{\Lambda^{f(\mathcal{R})}_m}{3}\,r^2+\frac{4\,M}{\sqrt{-\frac{\Lambda^{f(\mathcal{R})}_m}{3}}\,r}\right)}+r^2\,d\varphi^2-\frac{\Lambda^{f(\mathcal{R})}_m}{3}\,r^2\,dz^2,\label{BB16}
\end{eqnarray}
where $\Lambda^{f(\mathcal{R})}_m<0$ is given in Eq. (\ref{CC7}).

Now, we consider another function $f(\mathcal{R})$ given by the following form:
\begin{equation}\label{KK1}
    f(\mathcal{R})=\mathcal{R}+\alpha_k\,\mathcal{R}^{k+1},\quad k=1,2,....n.
\end{equation}

Substituting this function (\ref{KK1}), the Ricci tensor and the Ricci scalar provided in Eq. (\ref{a7}) into the modified field equations (\ref{CC2}), we obtain
\begin{equation}
    \mathcal{T}^{\mu\nu}=\left\{\Lambda_m+3\,\alpha^2+3\,\alpha^2\,(1-k)\,\alpha_k\,(-12\,\alpha^2)^{k}\right\}\,g^{\mu\nu}.\label{KK2}
\end{equation}

Solving the above field equations for zero energy-momentum tensor, $\mathcal{T}^{\mu\nu}=0$, and after simplification, we obtain 
\begin{equation}
    \Lambda_m=-3\,\alpha^2-3\,\alpha^2\,(1-k)\,\alpha_k\,(-12\,\alpha^2)^{k}.\label{KK3}
\end{equation}
In terms of the usual cosmological constant $\Lambda$, we can rewrite the effective cosmological constant as follows:
\begin{equation}
    \Lambda_m=\Lambda+\alpha_k\,(1-k)\,4^k\,\Lambda^{k+1},\quad k=1,2,....n.\label{KK4}
\end{equation}

The analysis in this section illustrates that the black hole solution represented by the line element (\ref{a1}) remains a valid solution within the $f(\mathcal{R})$-gravity framework. This solution describes a vacuum space-time characterized by an effective cosmological constant $\Lambda^{f(\mathcal{R})}_m$, as shown in Eq. (\ref{CC7}) as well as in Eq. (\ref{KK4}). The effective cosmological constant is influenced by the coupling constants: $\alpha_2$ associated with the $\mathcal{R}^3$ term, $\alpha_3$ and $\alpha_4$ related to the terms $\mathcal{R}^4$ and $\mathcal{R}^5$, respectively. Importantly, the expression for $\Lambda^{f(\mathcal{R})}_m$ indicates that the $\mathcal{R}^2$ term does not contribute to the effective cosmological constant; rather, it is the higher-order terms that play a significant role.

Below, we will explore BH space-time described in equation (\ref{a1}) within the framework of another modified gravity known as Ricci-Inverse ($\mathcal{RI}$) gravity. We will solve the modified field equations and discuss the resulting implications.

\section{Lemos BH solution within $\mathcal{RI}$-gravity framework}\label{sec:3}

In this section, we aim to examine the black hole solution described by equation (\ref{a1}) within the context of Ricci-Inverse gravity \cite{ref1} of Class-{\bf I} models, as suggested in Ref. \cite{ref2}. It is essential that the determinant of the Ricci tensor is non-zero for Ricci-Inverse gravity. In our current scenario, we find that the determinant of the Ricci tensor $R_{\mu\nu}$ presented in Eq. (\ref{a3}) is indeed non-zero, as given by
\begin{eqnarray}
    \mbox{det}\,(R_{\mu\nu})=-81\,\alpha^{10}\,r^4 \neq 0.\label{a10}
\end{eqnarray}
This non-zero determinant of the Ricci tensor indicates that there must existence an anti-curvature tensor $A^{\mu\nu}$ defined by
\begin{eqnarray}
    A^{\mu\nu}=R^{-1}_{\mu\nu}=\frac{\mbox{adj}\,(R_{\mu\nu})}{\mbox{det}\,(R_{\mu\nu})}.\label{b1}
\end{eqnarray}

Using the Ricci tensor given in equation (\ref{a3}), we derive the anti-curvature tensor $A^{\mu\nu}$ and its covariant form $A_{\mu\nu}$ as follows: 
\begin{eqnarray}
&&A^{\mu\nu}=\left(\begin{array}{cccc}
    \frac{r}{3\,\alpha\,(\alpha^3\,r^3-4\,M)} & 0 & 0 & 0 \\
     0 & \frac{4\,M}{3\,\alpha^3\,r}-\frac{r}{3} & 0 & 0\\
     0 & 0 & -\frac{1}{3\,\alpha^2\,r^2} & 0\\
     0 & 0 & 0 & -\frac{1}{3\,\alpha^4\,r^2}
\end{array}\right),\nonumber\\
&&A_{\mu\nu}=\left(\begin{array}{cccc}
    \frac{1}{3\,r}\,\left(r^3-\frac{4\,M}{\alpha^3}\right) & 0 & 0 & 0 \\
     0 & \frac{r}{3\,\alpha\,(4\,M-\alpha^3\,r^3)} & 0 & 0\\
     0 & 0 & -\frac{r^2}{3\,\alpha^2} & 0\\
     0 & 0 & 0 & -\frac{r^2}{3}
\end{array}\right).\label{b2}
\end{eqnarray}
And finally the anti-curvature scalar is given by
\begin{equation}
    \mathcal{A}=g_{\mu\nu}\,A^{\mu\nu}=-\frac{4}{3\,\alpha^2}=\frac{4}{\Lambda} \neq \frac{1}{\mathcal{R}}.\label{b3}
\end{equation}
From the above, we observe that the anti-curvature scalar $\mathcal{A}$ and the Ricci scalar $\mathcal{R}$ are not the same.

\subsection{BH solution in $\mathcal{RI}$-gravity: Class-{\bf I} model}\label{sec:3.1}

Now, we introduce anti-curvature tensor $A^{\mu\nu}$ into the Lagrangian of the system. Therefore, the action that describes the Ricci-Inverse gravity of Class-{\bf I} model with the function $f(\mathcal{R} ,\mathcal{A})=(\mathcal{R} + \beta\,\mathcal{A})$ is given by \cite{ref1,ref2}
\begin{eqnarray}
    \mathcal{S}= \int d^4x\, \sqrt{-g}\,\left[f(\mathcal{R} ,\mathcal{A})-2\,\Lambda_m+{\cal L}_m\right],\label{b4}
\end{eqnarray}
where $\beta$ is the coupling constant, $\Lambda_m$ is the cosmological constant in this modified gravity theory, ${\cal L}_m$ is the Lagrangian of the matter content. This action can be written as
\begin{eqnarray}
    \mathcal{S}=\int d^4x\, \sqrt{-g}\,\left[g_{\mu\nu}\,R^{\mu\nu} + \beta\,g_{\mu\nu}\,A^{\mu\nu}-2\,\Lambda_m+{\cal L}_m\right].\label{b5}
\end{eqnarray}

Now, varying the action (\ref{b5}) with respect to the metric tensor $g_{\mu\nu}$, the modified field equations that describe Class-{\bf I} model of Ricci-Inverse gravity theory is given by
\begin{eqnarray}
    R^{\mu\nu}-\frac{1}{2}\,\mathcal{R}\,g^{\mu\nu}+\Lambda_m\,g^{\mu\nu}+M^{\mu\nu}=\mathcal{T}^{\mu\nu},\label{b6}
\end{eqnarray}
with $\mathcal{T}^{\mu\nu}$ being the standard energy-momentum tensor and the tensor $M^{\mu \nu}$ is defined as
\begin{eqnarray}
     M^{\mu\nu}=-\beta\,A^{\mu\nu}-\frac{\mathcal{A}}{2}\,\beta\,g^{\mu\nu}+\frac{1}{2}\,\beta\,\Big[2\,g^{\rho\mu}\nabla_{\iota}\,\nabla_{\rho} (A^{\iota}_{\sigma}\,A^{\nu\sigma})-\nabla^2(A^{\mu}_{\iota}\,A^{\nu\iota})-g^{\mu\nu}\,\nabla_{\rho}\,\nabla_{\iota}(A^{\rho}_{\sigma}\,A^{\iota\sigma})\Big].\label{b7}
\end{eqnarray} 
Here, we have used $A^{\tau}_{\sigma}\,A^{\nu\sigma}=A^{\tau\lambda}\,g_{\lambda\sigma}\,A^{\nu\sigma}=A^{\tau\lambda}\,A^{\nu}_{\lambda}$.

Using the Ricci tensor and the Ricci scalar from Eq. (\ref{a7}), we can rewrite the modified field equations (\ref{b6}) as follows:
\begin{eqnarray}
    (-\Lambda+\Lambda_m)\,g^{\mu\nu}+M^{\mu\nu}=\mathcal{T}^{\mu\nu}.\label{b8}
\end{eqnarray}

Now, using the anti-curvature tensor presented in equation (\ref{b2}), we calculate the tensor $M^{\mu\nu}$ and its covariant form $M_{\mu\nu}$, which are given by: 
\begin{eqnarray}
&&M^{\mu\nu}=\left(\begin{array}{cccc}
     \frac{\beta\,r}{\alpha\,(4\,M-\alpha^3\,r^3)}& 0 & 0 & 0\\
     0 & \frac{\beta}{r}\,\left(r^3-\frac{4\,M}{\alpha^3}\right) & 0 & 0\\
     0 & 0 & \frac{\beta}{\alpha^2\,r^2} & 0\\
     0 & 0 & 0 & \frac{\beta}{\alpha^4\,r^2}
\end{array}\right),\nonumber\\
&&M_{\mu\nu}=\left(\begin{array}{cccc}
     \frac{\beta}{r}\,\left(\frac{4\,M}{\alpha^3}-r^3\right)& 0 & 0 & 0\\
     0 & \frac{\beta\,r}{\alpha\,(\alpha^3\,r^3-4\,M)} & 0 & 0\\
     0 & 0 & \frac{\beta\,r^2}{\alpha^2} & 0\\
     0 & 0 & 0 & \beta\,r^2
\end{array}\right).\label{b9}
\end{eqnarray}

Substituting the metric tensor $g^{\mu\nu}$ from equation (\ref{a2}) and the tensor $M^{\mu\nu}$ from (\ref{b9}) into the modified field equations (\ref{b8}), we obtain
\begin{eqnarray}
    \mathcal{T}^{\mu\nu}=\left(\Lambda_m+3\,\alpha^2+\frac{\beta}{\alpha^2}\right)\,\begin{pmatrix}
        -\left(\alpha^2\,r^2-\frac{4\,M}{\alpha\,r}\right)^{-1} & 0 & 0 & 0\\
        0 & \left(\alpha^2\,r^2-\frac{4\,M}{\alpha\,r}\right) & 0 & 0\\
        0 & 0 & \frac{1}{r^2} & 0\\
        0 & 0 & 0 & \frac{1}{\alpha^2,r^2}
    \end{pmatrix}\,.\label{b10}
\end{eqnarray}

Considering vacuum as the matter-energy content of the modified field equations whose energy-momentum tensor is $\mathcal{T}^{\mu\nu}=0$. Thereby, solving the modified field equations (\ref{b10}) in vacuum case and after simplification, we obtain  
\begin{eqnarray}
    \Lambda_m=-3\,\alpha^2-\frac{\beta}{\alpha^2}.\label{b111}
\end{eqnarray}
It is evident that when the coupling constant $\beta$ approaches zero ($\beta \to 0$), the effective cosmological constant simplifies to the usual one in GR, $\Lambda=-3\,\alpha^2$. Since $A^{\mu\nu}=R^{-1}_{\mu\nu}$, and thus, the anti-curvature tensor $A^{\mu\nu}$ and the anti-curvature scalar $\mathcal{A}$ have the dimension $[L^2]$. Therefore, the dimension of the coupling constant $\beta$ has $[L^{-4}]$ in the natural units.

Thus, we can conclude that the BH solution describe by the line-element (\ref{a1}) serves as a valid vacuum solution in Ricci-Inverse gravity theory of Class-{\bf I} model defined by the function $f(\mathcal{R} ,\mathcal{A})=(\mathcal{R} + \beta\,\mathcal{A})$ with a negative effective cosmological constant presented in Eq. (\ref{b111}). 

\subsection{BH solution in $\mathcal{RI}$-gravity: Class-{\bf II} model}\label{sec:3.2}

In this section, we will examine the same black hole solution described by equation (\ref{a1}) within the framework of Ricci-Inverse gravity \cite{ref1} of Class-{\bf II} models, as suggested in Ref. \cite{ref2}. For this model, the arbitrary function is defined as  $f=f(\mathcal{R},A^{\mu\nu}\,A_{\mu\nu})$. Therefore, the action that describes this $\mathcal{RI}$ gravitational theory of Class-{\bf II} models is given by:
\begin{eqnarray}
    \mathcal{S}=\int d^4x\, \sqrt{-g}\left[f(\mathcal{R}, A^{\mu\nu}\,A_{\mu\nu})-2\,\Lambda_m+{\cal L}_m\right].\label{A1}
\end{eqnarray}

To obtain the modified field equations, we perform a variation of the action (\ref{A1}) with respect to the metric tensor $g_{\mu\nu}$. This yields the following field equations for the Class-{\bf II} model:
\begin{eqnarray}
   \nonumber &&  f_{\mathcal{R}}\,R^{\mu \nu} - \frac{1}{2}\,f\,g^{\mu \nu}+\Lambda_m\, g^{\mu\nu} -  2\,f_{A^2}\,A^{\rho \nu}\,A^{\mu}_{\rho}     
 -  \nabla ^{\mu}\, \nabla ^{\nu}\,f_{\mathcal{R}} +  g^{\mu \nu}\, \nabla ^{\lambda}\,\nabla _{\lambda}\,f_{\mathcal{R}}  \nonumber\\
 &+& g^{\rho \nu}\, \nabla _{\alpha}\,  \nabla _{\rho}\,( f_{A^2}\,A_{\sigma \kappa}\, A^{\sigma \alpha}\,A^{\kappa \mu})    
 - \nabla ^2\,(f_{A^2}\,A_{\sigma \kappa}\, A^{\sigma \mu }\,A^{\kappa \nu}) - g^{\mu \nu}\, \nabla _{\alpha}\, \nabla _{\rho}\,(f_{A^2}\,A_{\sigma \kappa}\, A^{\sigma \alpha}\,A^{\kappa \rho})    \nonumber\\ 
     &+& 2\,g^{\rho \nu}\, \nabla _{\rho} \nabla _{\alpha}( f_{A^2}\,A_{\sigma \kappa}\, A^{\sigma \mu }\,A^{\kappa \alpha} )    -g^{\rho \nu}\, \nabla _{\alpha} \nabla _{\rho}\,(f_{A^2}\,A_{\sigma \kappa}\, A^{\sigma \mu}\,A^{\kappa \alpha})     =\mathcal{T}^{\mu\nu},\label{A2}
\end{eqnarray}
where $\mathcal{T}^{\mu\nu}$ being the standard energy-momentum tensor and
\begin{equation}
    f_{\mathcal{R}}=\partial f/\partial \mathcal{R},\quad\quad f_{A^2}=\partial f/\partial (A^{\mu\nu}\,A_{\mu\nu}).\label{AA}
\end{equation}

In order to express the modified field equations (\ref{A2}) more succinctly, let us define the following tensors
\begin{eqnarray}
Y^{\mu \nu}\equiv 2\,f_{A^2}\,A^{\rho \nu}\,A^{\mu}_{\rho}\,. \label{A3}
\end{eqnarray}
And 
\begin{eqnarray}
    \nonumber U^{\mu \nu}&\equiv& - \nabla ^{\mu} \nabla ^{\nu}f_{\mathcal{R}} +  g^{\mu \nu}\, \nabla ^{\lambda}\,\nabla _{\lambda}f_{\mathcal{R}}        + g^{\rho \nu}\, \nabla _{\alpha}  \nabla _{\rho}( f_{A^2}\,A_{\sigma \kappa}\, A^{\sigma \alpha}\,A^{\kappa \mu})    \\
   \nonumber &-& \nabla ^2(f_{A^2}\,A_{\sigma \kappa}\, A^{\sigma \mu }\,A^{\kappa \nu}) - g^{\mu \nu}\, \nabla _{\alpha} \nabla _{\rho}(f_{A^2}\,A_{\sigma \kappa}\, A^{\sigma \alpha}\,A^{\kappa \rho})      \\ 
     &+& 2\,g^{\rho \nu}\, \nabla _{\rho} \nabla _{\alpha}( f_{A^2}\,A_{\sigma \kappa}\, A^{\sigma \mu }\,A^{\kappa \alpha})-g^{\rho \nu}\, \nabla _{\alpha} \nabla _{\rho}(f_{A^2}\,A_{\sigma \kappa}\,A^{\sigma \mu}\,A^{\kappa \alpha}) \, .\label{A4}
\end{eqnarray}

Therefore, the modified field equation (\ref{A2}) can be rewritten as follows:
\begin{eqnarray}\label{A5}
  f_{\mathcal{R}}\, R^{\mu \nu} - \frac{1}{2}\,f\,g^{\mu \nu}+\Lambda_m\, g^{\mu\nu} -  Y^{\mu \nu}  +U^{\mu \nu} =\mathcal{T}^{\mu \nu}.
\end{eqnarray}

In Class-{\bf II} model, we consider the function $f(\mathcal{R}, A^{\mu\nu}\,A_{\mu\nu})$ to be the following form:
\begin{equation} \label{A6}
    f(\mathcal{R}, A^{\mu \nu}\,A_{\mu \nu}) =\mathcal{R}+\gamma\, A^{\mu \nu}\,A_{\mu \nu}, 
\end{equation}
where $\gamma$ being an arbitrary constant. This function leads to the following non-zero derivative
\begin{eqnarray}\label{A7}
f_{\mathcal{R}}=1,\quad\quad f_{A^2}=\gamma.
\end{eqnarray}

It is important to note that one should consider higher-order terms of both the Ricci scalar $\mathcal{R}$ and the quadratic invariant $A^{\mu \nu}\,A_{\mu \nu}$ in the function $f(\mathcal{R}, A^{\mu \nu}\,A_{\mu \nu})$. For example, we can express it as $f(\mathcal{R}, A^{\mu \nu}\,A_{\mu \nu})=\mathcal{R}+\sum_{i}\,\alpha_i\,\mathcal{R}^{1+i}+\sum_{i}\,\gamma_i\, (A^{\mu \nu}\,A_{\mu \nu})^i$, where $i=1,2,....n$.

Substituting the function $f$ from Eq. (\ref{A6}) into the modified Eq. (\ref{A5}) results
\begin{equation}\label{A8}
    R^{\mu\nu}-\frac{1}{2}\,\mathcal{R}\,g^{\mu\nu}-\frac{\gamma}{2}\,A^{\mu\nu}\,A_{\mu\nu}\,g^{\mu\nu}-2\,\gamma\,A^{\rho\nu}\,A^{\mu}_{\rho}+U^{\mu\nu}+\Lambda_{m}\,g^{\mu\nu}=\mathcal{T}^{\mu\nu},
\end{equation}
where
\begin{eqnarray}\label{A9}
    U^{\mu\nu}&=&\gamma\,\Big[g^{\rho \nu} \nabla _{\alpha}  \nabla _{\rho}(A_{\sigma \kappa} A^{\sigma \alpha}A^{\kappa \mu})-\nabla ^2(A_{\sigma \kappa} A^{\sigma \mu }A^{\kappa \nu})-g^{\mu \nu} \nabla _{\alpha} \nabla _{\rho}(A_{\sigma \kappa} A^{\sigma \alpha}A^{\kappa \rho}) \nonumber\\
   &+&2\,g^{\rho \nu} \nabla _{\rho} \nabla _{\alpha}(A_{\sigma \kappa} A^{\sigma \mu }A^{\kappa \alpha})-g^{\rho \nu} \nabla _{\alpha} \nabla _{\rho}(A_{\sigma \kappa} A^{\sigma \mu}A^{\kappa \alpha})\Big]\,.
\end{eqnarray}

Thus, by substituting the metric tensor from equation (\ref{a2}) and the anti-curvature tensor from equation (\ref{b2}) into the modified field equations (\ref{A8}), we obtain:
\begin{eqnarray}\label{A10}
    \mathcal{T}^{\mu\nu}=\left(\Lambda_m+3\,\alpha^2-\frac{4\,\gamma}{9\,\alpha^4}\right)\,\begin{pmatrix}
        -\left(\alpha^2\,r^2-\frac{4\,M}{\alpha\,r} \right)^{-1} & 0 & 0 & 0\\
        0 & \left(\alpha^2\,r^2-\frac{4\,M}{\alpha\,r} \right) & 0 & 0\\
        0 & 0 & \frac{1}{r^2} & 0\\
        0 & 0 & 0 & \frac{1}{\alpha^2\,r^2}
    \end{pmatrix}\,.
\end{eqnarray}

To solve the modified field equations (\ref{A10}) in a vacuum where the matter-energy content is represented by an energy-momentum tensor of $\mathcal{T}^{\mu\nu}=0$, we can simplify Eq. (\ref{A10})
\begin{equation}
    \Lambda_m=-3\,\alpha^2+\frac{4\,\gamma}{9\,\alpha^4}\,.\label{A11}
\end{equation}

From the analysis above, it is evident that as the coupling constant approaches zero, $\gamma \to 0$, the effective cosmological constant reduces to the usual general relativity one, given by $\Lambda=-3\,\alpha^2$. Given that the usual cosmological constant $\Lambda$ is negative, the effective cosmological constant $\Lambda_m$ will also be negative, provided the following condition is satisfied:
\begin{itemize}
    \item $\Lambda <0$,\quad $\gamma>0$, \quad $\Lambda_m<0$\quad provided\quad $\gamma<-\Lambda^3/4$.
    \item $\Lambda <0$,\quad $\gamma<0$, \quad $\Lambda_m<0$.
\end{itemize}

Thus, we observe that the black hole solution described by the line element (\ref{a1}) is a valid vacuum solution in the Ricci-Inverse gravity framework of the Class-{\bf II} model, with the function $f(\mathcal{R}, A^{\mu \nu}\,A_{\mu \nu})=\mathcal{R}+\gamma\, A^{\mu \nu}\,A_{\mu \nu}$. Noted that $A^{\mu\nu}=R^{-1}_{\mu\nu}$, and thus, the anti-curvature tensor $A^{\mu\nu}$ and the anti-curvature scalar $\mathcal{A}$ have the dimension $[L^2]$. Therefore, the dimension of the coupling constant $\gamma$ has $[L^{-6}]$ in the natural units.

\subsection{BH solution with $\mathcal{RI}$-gravity: Class-{\bf III} model}\label{sec:3.3}

In this section, we explore this BH space-time described by the line-element (\ref{a1}) in Class-{\bf III} model of Ricci-Inverse gravity. In this Class-{\bf III} model, the function $f=f(\mathcal{R},\mathcal{A},A^{\mu \nu}A_{\mu \nu})$. Therefore, the action describing Class-{\bf III} model of $\mathcal{RI}$-gravity is given by
\begin{equation}\label{B1}
    \mathcal{S}=\int \mathrm{d}x^4 \sqrt{-g}\,[f(\mathcal{R},\mathcal{A},A^{\mu \nu}A_{\mu \nu})-2\,\Lambda_m]+ S_m.
\end{equation}

By varying the action (\ref{B1}) with respect to the metric tensor $g_{\mu\nu}$, the modified field equations in Class-{\bf III} model of Ricci-Inverse gravity are given by  
\begin{equation}\label{B2}
    -\frac{1}{2}\,f\,g^{\mu\nu}+f_{\mathcal{R}}\,R^{\mu \nu}-f_{\mathcal{A}}\,A^{\mu\nu}-2\,f_{A^2}\,A^{\rho \nu }\,A^{\mu}_{\rho}+P^{\mu\nu}+M^{\mu\nu}+U^{\mu\nu}+ \Lambda_m\,g^{\mu\nu}=\mathcal{T}^{\mu\nu},
\end{equation}
where 
\begin{eqnarray}
    P^{\mu\nu}&=&g^{\mu\nu}\,\nabla ^2\,f_{\mathcal{R}}-\nabla^{\mu} \nabla^{\nu}\,f_{\mathcal{R}},\label{B3}\\ 
    M^{\mu\nu}&=&g^{\rho\mu}\,\nabla_{\alpha}\,\nabla_{\rho }(f_{\mathcal{A}}\, A_{\sigma}^{\alpha}\,A^{\nu\sigma})-\frac{1}{2}\,\nabla^2 (f_{\mathcal{A}}\,A^{\mu}_{\sigma}\,A^{\nu\sigma})-\frac{1}{2}\,g^{\mu\nu}\, \nabla_{\alpha}\,\nabla_{ \rho}(f_{\mathcal{A}}\,A_{\sigma}^{\alpha}\,A^{\rho \sigma}),\label{B4}\\
 \nonumber  U^{\mu\nu}&=&g^{\rho\nu}\,\nabla_{\alpha}\,\nabla_{\rho}(f_{A^2}\,A_{\sigma\kappa}\,A^{\sigma\alpha}\,A^{\mu\kappa})-\nabla^{2}(f_{A^2}\,A_{\sigma \kappa}\,A^{\sigma\mu}\,A^{\nu\kappa})
  -g^{\mu\nu}\,\nabla_{\alpha}\,\nabla_{\rho}(f_{A^2}\,A_{\sigma \kappa}\,A^{\sigma\alpha}\,A^{\rho\kappa})\nonumber\\
  &+&2\,g^{\rho\nu}\,\nabla_{\rho}\,\nabla_{\alpha}(f_{A^2}\,A_{\sigma \kappa}\,A^{\sigma\mu}\,A^{\alpha\kappa})
   - g^{\rho\nu}\,\nabla_{\alpha}\,\nabla_{\rho}(f_{A^2}\,A_{\sigma \kappa}\,A^{\sigma\mu}\,A^{\alpha\kappa}).\label{B5} 
\end{eqnarray}
Here $f_{\mathcal{R}},f_{\mathcal{A}}$ and $f_{A^2}=\partial f/\partial (A^{\mu\nu}\,A_{\mu\nu})$ are defined earlier.

Let's consider the function $f$ in this Class-{\bf III} model to be the following form:
\begin{equation} \label{B7}
    f(\mathcal{R},\mathcal{A}, A^{\mu \nu}\,A_{\mu \nu}) =\mathcal{R}+\alpha_1\,\mathcal{R}^2+\alpha_2\,\mathcal{R}^3+\beta_1 \, \mathcal{A} +\beta_2 \, \mathcal{A}^2+\gamma\,A^{\mu \nu}\,A_{\mu \nu} \, 
\end{equation}
with $\alpha_i,\beta_i$ and $\gamma$ being arbitrary constants. Since $A^{\mu\nu}=R^{-1}_{\mu\nu}$, and thus, the anti-curvature tensor $A^{\mu\nu}$ and the anti-curvature scalar $\mathcal{A}$ have the dimension $[L^2]$. Therefore, the dimensions of the various coupling constants are as follows: $\beta_1$ has dimension $[L^{-4}]$, $\beta_2$ has dimension $[L^{-6}]$, and $\gamma$ has dimension $[L^{-6}]$ in the natural units.

Using the above function (\ref{B7}), one can find the following
\begin{eqnarray}\label{B8}
f_{\mathcal{R}}=1+2\,\alpha_1\,\mathcal{R}+3\,\alpha_2\,\mathcal{R}^2,\quad\quad    f_{\mathcal{A}}=\beta_1+2\,\beta_2\,\mathcal{A},\quad\quad 
    f_{A^2}=\gamma.
\end{eqnarray}

Therefore, the modified field equations (\ref{B2}) using (\ref{B7})--(\ref{B8}) reduces to the following form:
\begin{eqnarray}\label{B9}
    &&-\frac{1}{2}\,(\mathcal{R}+\alpha_1\,\mathcal{R}^2+\alpha_2\,\mathcal{R}^3+\beta_1\,\mathcal{A}+\beta_2\,\mathcal{A}^2+\gamma\,A^{\mu\nu}\,A_{\mu\nu})\,g^{\mu\nu} +(1+2\,\alpha_1\,\mathcal{R}+3\,\alpha_2\,\mathcal{R}^2)\,R^{\mu\nu}\nonumber\\
    &&-(\beta_1+2\,\beta_2\,\mathcal{A})\,A^{\mu\nu}-2\,\gamma\,A^{\rho\nu }\,A^{\mu}_{\rho}+M^{\mu\nu}+U^{\mu\nu}+\Lambda_m\,g^{\mu\nu}=\mathcal{T}^{\mu\nu},
\end{eqnarray}
where 
\begin{eqnarray}
&&P^{\mu \nu}=g^{\mu\nu}\,\nabla^2(1+2\,\alpha_1\,\mathcal{R}++3\,\alpha_2\,\mathcal{R}^2)-\nabla^{\mu}\,\nabla^{\nu}(1+2\,\alpha_1\,\mathcal{R}++3\,\alpha_2\,\mathcal{R}^2)=0,\nonumber\\
&&M^{\mu\nu}=g^{\rho\mu}\,\nabla_{\tau}\,\nabla_{\rho} [(\beta_1+2\,\beta_2\,\mathcal{A})\,A_{\sigma}^{\tau}\,A^{\nu \sigma}]-\frac{1}{2}\nabla^2\, [(\beta_1+2\,\beta_2\,\mathcal{A})\,A^{\mu}_{\sigma}\,A^{\nu \sigma}]\nonumber\\
&&-\frac{1}{2}\,g^{\mu\nu}\,\nabla_{\tau}\,\nabla_{\rho}\,[ (\beta_1+2\,\beta_2\,\mathcal{A})\,A_{\sigma}^{\tau}\,A^{\rho \sigma}]\nonumber\\
&&=(\beta_1+2\,\beta_2\,\mathcal{A})\,\Big\{g^{\rho\mu}\,\nabla_{\tau}\,\nabla_{\rho} (A_{\sigma}^{\tau}\,A^{\nu\sigma})-\frac{1}{2}\,\nabla^2\,(A^{\mu}_{\sigma}\,A^{\nu \sigma})-\frac{1}{2}\,g^{\mu\nu}\,\nabla_{\tau}\,\nabla_{\rho}\,(A_{\sigma}^{\tau}\,A^{\rho \sigma})\Big\},\nonumber\\ 
&&U^{\mu\nu}=\gamma\,\Big[g^{\rho\nu}\,\nabla_{\tau}\,\nabla_{\rho}(A_{\sigma\kappa}\,A^{\sigma\alpha}\,A^{\mu\kappa})-\nabla^{2}\,(A_{\sigma\kappa}\,A^{\sigma\mu}\,A^{\nu\kappa})-g^{\mu \nu}\,\nabla_{\tau}\,\nabla_{\rho}\,(A_{\sigma\kappa}\,A^{\sigma \tau}\,A^{\rho\kappa})\nonumber\\
  &+&2\,g^{\rho\nu}\,\nabla_{\rho}\,\nabla_{\tau}(A_{\sigma \kappa}\,A^{\sigma\mu}\,A^{\tau\kappa})-g^{\rho\nu}\,\nabla _{\tau}\,\nabla_{\rho}(A_{\sigma\kappa}\,A^{\sigma \mu}\,A^{\tau \kappa})\Big].\label{B11} 
\end{eqnarray}

Substituting the Ricci tensor $R_{\mu\nu}$ and the Ricci scalar $\mathcal{R}$ from Eq.(\ref{a7}), and the anti-curvature tensor $A^{\mu\nu}$ from Eq. (\ref{b2}) into the modified fields equations (\ref{B9}), we obtain
\begin{eqnarray}\label{B12}
    &&\mathcal{T}^{\mu\nu}=\tilde{\Lambda}\,
    \begin{pmatrix}
        -\left(\alpha^2\,r^2-\frac{4\,M}{\alpha\,r} \right)^{-1} & 0 & 0 & 0\\
        0 & \left(\alpha^2\,r^2-\frac{4\,M}{\alpha\,r} \right) & 0 & 0\\
        0 & 0 & \frac{1}{r^2} & 0\\
        0 & 0 & 0 & \frac{1}{\alpha^2\,r^2}
    \end{pmatrix}=\tilde{\Lambda}\,g^{\mu\nu},\nonumber\\
    &&\tilde{\Lambda}=\left(\Lambda_m+3\,\alpha^2+\frac{\beta_1}{\alpha^2}-\frac{16\,\beta_2}{9\,\alpha^4}-\frac{4\,\gamma}{9\,\alpha^4}-432\,\alpha^6\,\alpha_2\right).
\end{eqnarray}

Considering vacuum as the matter-energy content of the modified field equations whose energy-momentum tensor is $\mathcal{T}^{\mu\nu}=0$. Thereby, solving the modified field equations (\ref{B12}) in vacuum case and after simplification, we obtain  
\begin{equation}
    \Lambda^{\mathcal{RI}}_m=-3\,\alpha^2+432\,\alpha^6\,\alpha_2-\frac{\beta_1}{\alpha^2}+\frac{16\,\beta_2}{9\,\alpha^4}+\frac{4\,\gamma}{9\,\alpha^4}.\label{B13}
\end{equation}

In terms of usual cosmological constant $\Lambda$, we can write the effective cosmological constant as follows:
\begin{equation}
    \Lambda^{\mathcal{RI}}_m=\Lambda-16\,\Lambda^3\,\alpha_2+\frac{3\,\beta_1}{\Lambda}+\frac{16\,\beta_2}{\Lambda^2}+\frac{4\,\gamma}{\Lambda^2}.\label{B14}
\end{equation}

The analysis demonstrates that the black hole solution represented by the line-element (\ref{a1}) remains a valid solution within the Ricci-Inverse gravity framework of Class-{\bf III} model. This solution remains a vacuum space-time characterized by an effective cosmological constant $\Lambda_m$, as articulated in Eq. (\ref{B14}). This constant is influenced by several coupling constants: $\alpha_2$ related to the $\mathcal{R}^3$ term, $\beta_1$ and $\beta_2$ associated with the terms $\mathcal{A}$ and $\mathcal{A}^2$, respectively, and $\gamma$ linked to the quadratic invariant $A^{\mu\nu}\,A_{\mu\nu}$. Notably, the expression for $\Lambda_m$ shows that the $\mathcal{R}^2$ term does not contribute to the effective cosmological constant, underscoring a compelling feature of this model. 

The effective cosmological constant $\Lambda_m$ is negative provided the following condition must hold:
\begin{equation}\label{B17}
    \frac{16\,\beta_2}{\Lambda^2}+\frac{4\,\gamma}{\Lambda^2}-16\,\Lambda^3\,\alpha_2<-\left(\Lambda+\frac{3\,\beta_1}{\Lambda}\right).
\end{equation}

Moreover, the exploration of this novel Ricci-Inverse gravity theory can be significantly advanced by incorporating higher-order terms $\mathcal{R}^i$ for $i>3$ in the function $f(\mathcal{R},\mathcal{A},A^{\mu\nu}\,A_{\mu\nu})$. In the subsequent section, we will analyze the implications of this cylindrical space-time (\ref{a1}) within the $f(\mathcal{R})$-gravity framework. This examination aims to yield deeper insights into the characteristics of the space-time, enhancing our understanding of the intricate relationship between geometric terms and the underlying gravitational theories.

\begin{figure}[ht!]
    \centering
    \includegraphics[width=0.75\linewidth]{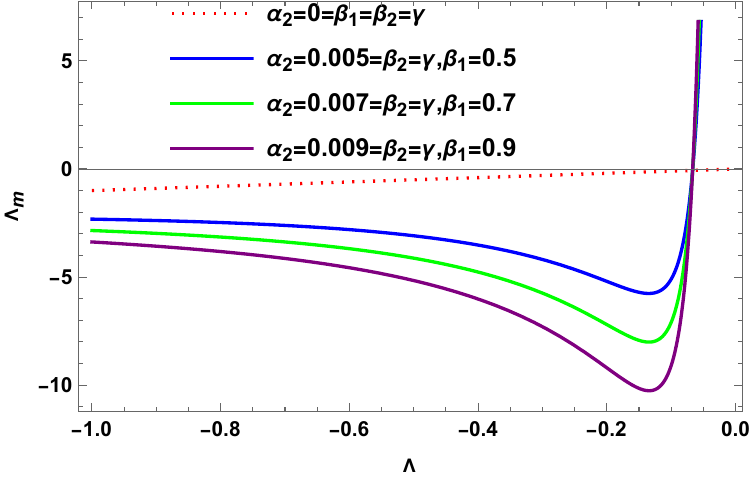}
    \caption{The effective cosmological constant $\Lambda_m$ as a function of the usual cosmological constant $\Lambda$.}
    \label{fig:1}
\end{figure}

In Figure \ref{fig:1}, we illustrate the effective cosmological constant $\Lambda_m$ as a function of the usual cosmological constant $\Lambda$. The figure showcases the variations of $\Lambda_m$ for different values of the coupling constants $\alpha_2,\beta_1,\beta_2$ and $\gamma$. The dotted red line represents the cases in general relativity, while the solid lines correspond to various values of the coupling constants. This visualization underscores the impact of these coupling constants on the relationship between the effective and traditional cosmological constants, offering valuable insights into the dynamics of the model.

The BH solution (\ref{a1}) in terms of usual cosmological constant ($\Lambda$) within the framework of general relativity is described by the following line-element
\begin{eqnarray}
ds^2=-\left(-\frac{\Lambda}{3}\,r^2-\frac{4\,M}{\sqrt{-\frac{\Lambda}{3}}\,r}\right)\,dt^2+\frac{dr^2}{\left(-\frac{\Lambda}{3}\,r^2-\frac{4\,M}{\sqrt{-\frac{\Lambda}{3}}\,r}\right)}+r^2\,d\varphi^2-\frac{\Lambda}{3}\,r^2\,dz^2.\label{B15}
\end{eqnarray}

The BH solution (\ref{a1}) in terms of effective cosmological constant $\Lambda_m$ within the framework of Ricci-Inverse gravity is therefore described by the following line-element
\begin{eqnarray}
ds^2=\left(\frac{\Lambda^{\mathcal{RI}}_m}{3}\,r^2+\frac{4\,M}{\sqrt{-\frac{\Lambda^{\mathcal{RI}}_m}{3}}\,r}\right)\,dt^2-\frac{dr^2}{\left(\frac{\Lambda^{\mathcal{RI}}_m}{3}\,r^2+\frac{4\,M}{\sqrt{-\frac{\Lambda^{\mathcal{RI}}_m}{3}}\,r}\right)}+r^2\,d\varphi^2-\frac{\Lambda^{\mathcal{RI}}_m}{3}\,r^2\,dz^2,\label{B16}
\end{eqnarray}
where $\Lambda^{\mathcal{RI}}_m<0$ is given in Eq. (\ref{B14}).

\section{Geodesic motions of test particles in $f(\mathcal{R})$ and $\mathcal{RI}$- gravity }\label{sec:4}

In this section, we examine the geodesic motions of test particles in the vicinity of the black hole solutions described in Equations (\ref{BB16}) and (\ref{B16}), which are derived from the Ricci-Inverse and $f(\mathcal{R})$-gravity theories, respectively, as discussed in the preceding sections. The Lagrangian of a system is defined by
\begin{equation}
    \mathcal{L}=\frac{1}{2}\,g_{\mu\nu}\,\dot{x}^{\mu}\,\dot{x}^{\nu},\label{C1}
\end{equation}
where dot represents ordinary derivative w. r. t. affine parameter $\lambda$.

For the space-time (\ref{BB16}) and (\ref{B16}), we obtain the following expression of the Lagrangian function given by:
\begin{equation}
    \mathcal{L}=\frac{1}{2}\,\Bigg[-\left(-\frac{\Lambda_m}{3}\,r^2-\frac{4\,M}{\sqrt{-\frac{\Lambda_m}{3}}\,r}\right)\,\dot{t}^2+\frac{\dot{r}^2}{\left(-\frac{\Lambda_m}{3}\,r^2-\frac{4\,M}{\sqrt{-\frac{\Lambda_m}{3}}\,r}  \right)}+r^2\,\dot{\varphi}^2+\frac{(-\Lambda_m)}{3}\,r^2\,\dot{z}^2\Bigg],\label{C2}
\end{equation}
where in $f(\mathcal{R})$-gravity framework, $\Lambda_m$ is defined by $\Lambda^{f(\mathcal{R})}_m$ as presented in equation (\ref{CC7}). Similarly, in Ricci-Inverse gravity, the parameter $\Lambda_m$ transform to $ \Lambda^{\mathcal{RI}}_m$ as outlined in equation (\ref{B14}). 

From the LBH metric (\ref{a1}), it is evident that the metric tensor $g_{\mu\nu}$ depends solely on the radial distance $r$ and is independent of the coordinates $(t, \varphi, z)$. Consequently, there are three conserved quantities, which are given by:
\begin{eqnarray}
    &&-\mathrm{E}=\frac{d\mathcal{L}}{d\dot{t}}=-g_{tt}\,\dot{t}\Rightarrow \dot{t}=\frac{\mathrm{E}}{g_{tt}},\quad\quad g_{tt}=\left(-\frac{\Lambda_m}{3}\,r^2-\frac{4\,M}{\sqrt{-\frac{\Lambda_m}{3}}\,r}\right),\label{C3}\\
    && \mathrm{L}=\frac{d\mathcal{L}}{d\dot{\varphi}}=r^2\,\dot{\varphi}\Rightarrow \dot{\varphi}=\frac{\mathrm{L}}{r^2},\label{C4}\\
    &&\mathrm{p}_z=\frac{d\mathcal{L}}{d\dot{z}}=g_{zz}\,\dot{z} \Rightarrow \dot{z}=\frac{\mathrm{p}_z}{g_{zz}},\quad\quad g_{zz}=\frac{(-\Lambda_m)}{3}\,r^2.\label{C5}
\end{eqnarray}

Using these conserved quantities, we can express the equations for time-like and null geodesics as follows:
\begin{equation}
    -\frac{\mathrm{E}^2}{\left(-\frac{\Lambda_m}{3}\,r^2-\frac{4\,M}{\sqrt{-\frac{\Lambda_m}{3}}\,r}\right)}+\frac{\dot{r}^2}{\left(-\frac{\Lambda_m}{3}\,r^2-\frac{4\,M}{\sqrt{-\frac{\Lambda_m}{3}}\,r}\right)}+\frac{\mathrm{L}^2}{r^2}+\frac{3}{(-\Lambda_m)}\,\frac{\mathrm{p}^2_z}{r^2}=\varepsilon,\label{C6}
\end{equation}
where $\varepsilon=0$ for null-geodesics and $-1$ for time-like.

\begin{center}
\begin{figure}[ht!]
\subfloat[$\alpha_2=0.1=\alpha_3,\alpha_4=0.001$]{\centering{}\includegraphics[scale=0.6]{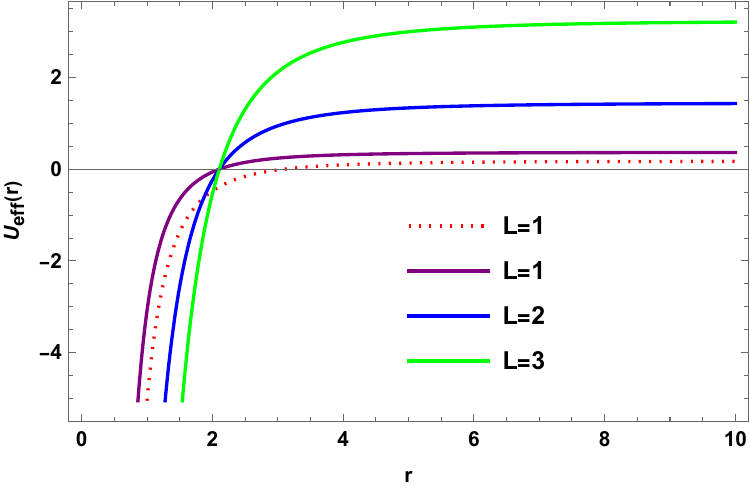}}\quad\quad
\subfloat[]{\centering{}\includegraphics[scale=0.6]{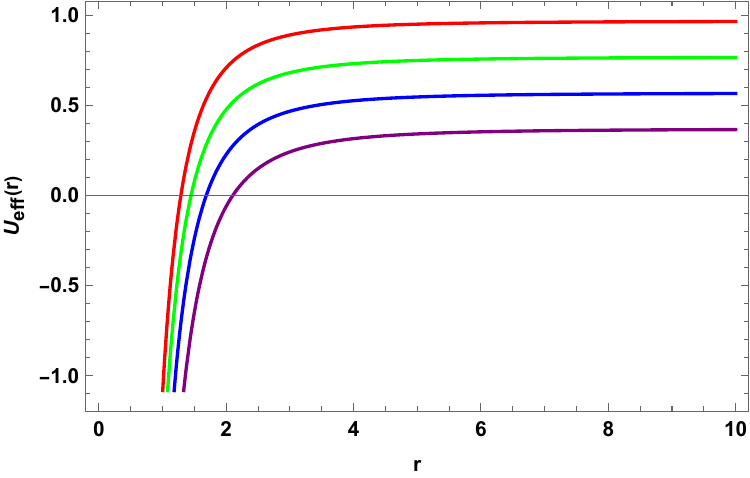}}
\centering{}\caption{The effective potential of null geodesics within $f(\mathcal{R})$-gravity. Here we set $M=1/2, \Lambda=-0.5, \mathrm{p}_z=0.1$.}\label{fig:5}
\end{figure}
\par\end{center}

The above equation (\ref{C6}) can be written as
\begin{equation}
    \left(\frac{dr}{d\lambda}\right)^2+\mathrm{U}_\text{eff} (r)=\mathrm{E}^2\,\label{C7}
\end{equation}
which is the one-dimensional equation of motion of a test particles having energy $\mathrm{E}^2$ and the effective potential $\mathrm{U}_{eff}(r)$ given by

\begin{center}
\begin{figure}[ht]
\subfloat[$\alpha_2=0.1=\alpha_3,\alpha_4=0.001$]{\centering{}\includegraphics[scale=0.6]{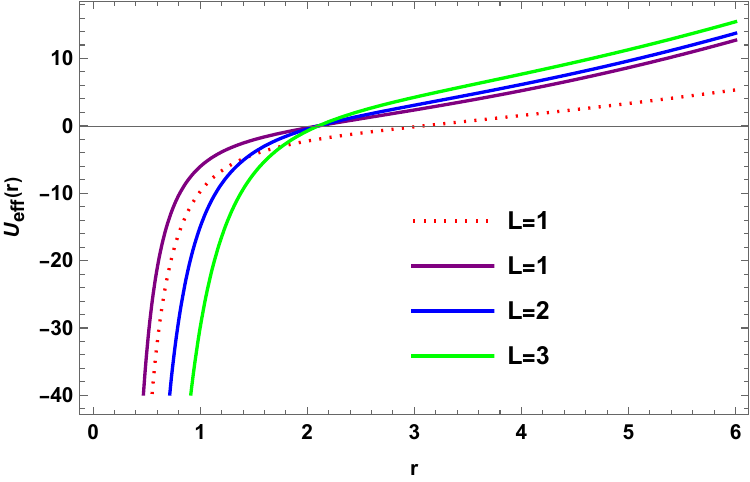}}\quad\quad
\subfloat[]{\centering{}\includegraphics[scale=0.6]{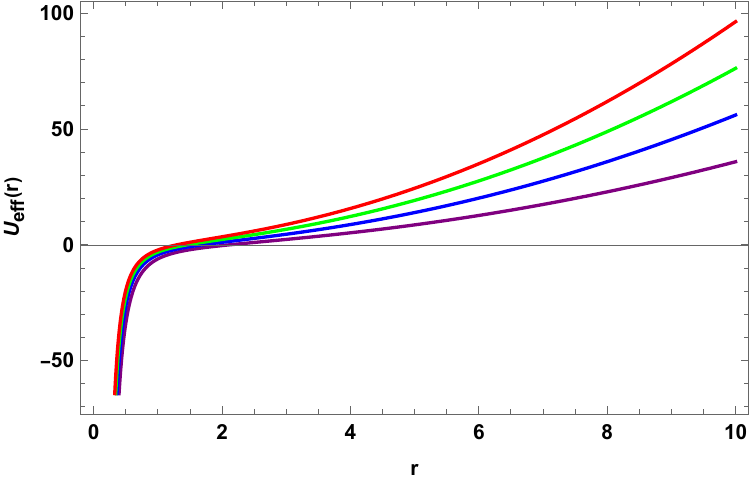}}
\centering{}\caption{The effective potential of time-like geodesics within $f(\mathcal{R})$-gravity. Here we set $M=1/2, \Lambda=-0.5, \mathrm{p}_z=0.1$.}\label{fig:6}
\end{figure}
\par\end{center}

\begin{small}
\begin{eqnarray}
    \mathrm{U}_\text{eff}&=&\left(-\frac{r^2}{3}\,\Lambda-\frac{4\,M}{r\,\sqrt{-\frac{\Lambda}{3}}}\right]\quad \mbox{in $\text{GR}$},\label{C888}\\
    \mathrm{U}^{f(\mathcal{R})}_\text{eff}&=&\left[-\frac{r^2}{3}\left(\Lambda-16\Lambda^3\alpha_2-128\Lambda^4\alpha_3-768\alpha_4\Lambda^5\right)-\frac{4M}{r\sqrt{-\frac{1}{3}\left(\Lambda-16\Lambda^3\alpha_2-128\Lambda^4\alpha_3-768\alpha_4\Lambda^5\right)}}\right]\times\nonumber\\
    &&\left[-\varepsilon+\frac{\mathrm{L}^2}{r^2}-\frac{3\mathrm{p}^2_z}{r^2\left(\Lambda-16\,\Lambda^3\,\alpha_2-128\,\Lambda^4\,\alpha_3-768\,\alpha_4\,\Lambda^5\right)}\right]\quad \mbox{in $f(\mathcal{R})$ gravity}.\label{C8}\\
    \mathrm{U}^{\mathcal{RI}}_\text{eff}&=&\left[-\frac{r^2}{3}\left(\Lambda+16\Lambda^3\alpha_2+\frac{3\beta_1}{\Lambda}+\frac{16\beta_2}{\Lambda^2}+\frac{4\gamma}{\Lambda^2}\right)-\frac{4M}{r\,\sqrt{-\frac{1}{3}\,\left(\Lambda+16\Lambda^3\alpha_2+\frac{3\beta_1}{\Lambda}+\frac{16\beta_2}{\Lambda^2}+\frac{4\gamma}{\Lambda^2}\right)}}\right]\times\nonumber\\
    &&\left[-\varepsilon+\frac{\mathrm{L}^2}{r^2}-\frac{3\mathrm{p}^2_z}{r^2\left(\Lambda+16\Lambda^3\alpha_2+\frac{3\beta_1}{\Lambda}+\frac{16\beta_2}{\Lambda^2}+\frac{4\gamma}{\Lambda^2}\right)}\right]\quad \mbox{in Ricci-Inverse gravity}.\label{C88}
\end{eqnarray}
\end{small}

The analysis demonstrates that the effective potential of the system is significantly shaped by the coupling constants present in both $f(\mathcal{R})$ and $\mathcal{RI}$-gravity theories. In $f(\mathcal{R})$-gravity framework, the relevant coupling constants are $\alpha_2,\alpha_3$, and $\alpha_4$, while in $\mathcal{RI}$-gravity, the key parameters are $\alpha_2,\beta_1,\beta_2$ and $\gamma$. This highlights the intricate interplay of these constants in determining the system's effective potential.

\begin{center}
\begin{figure}[ht!]
\subfloat[$\alpha_2=\beta_2=\gamma=0.005,\beta_1=0.5$]{\centering{}\includegraphics[scale=0.63]{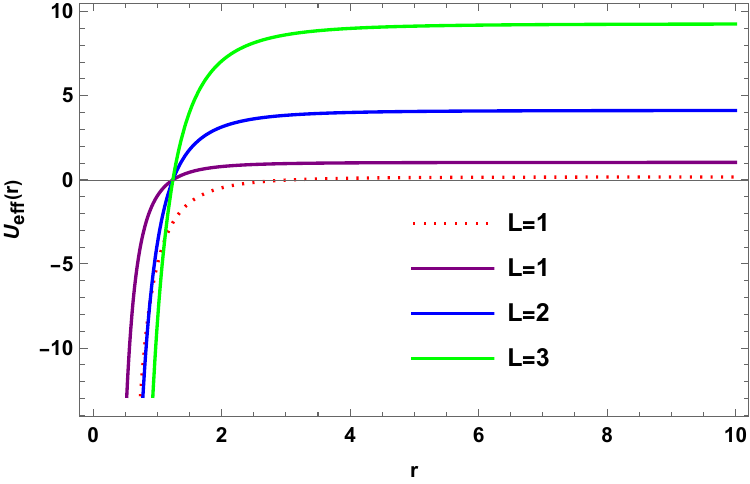}}\quad\quad
\subfloat[]{\centering{}\includegraphics[scale=0.63]{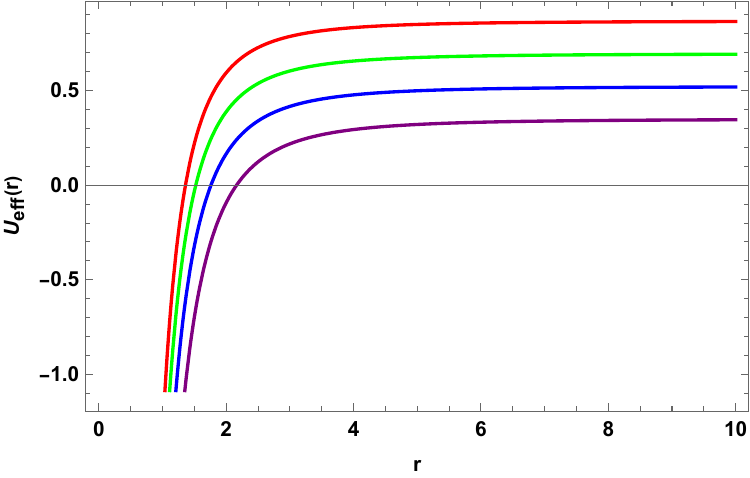}}
\centering{}\caption{The effective potential of null geodesics within $\mathcal{RI}$-gravity. Here we set $M=1/2, \Lambda=-0.5, \mathrm{p}_z=0.1$.}\label{fig:3}
\hfill\\
\subfloat[$\alpha_2=\beta_2=\gamma=0.005,\beta_1=0.5$]{\centering{}\includegraphics[scale=0.63]{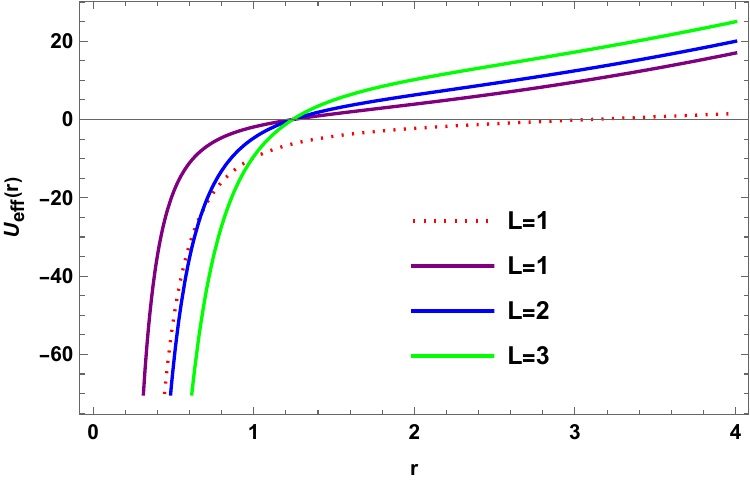}}\quad\quad
\subfloat[]{\centering{}\includegraphics[scale=0.63]{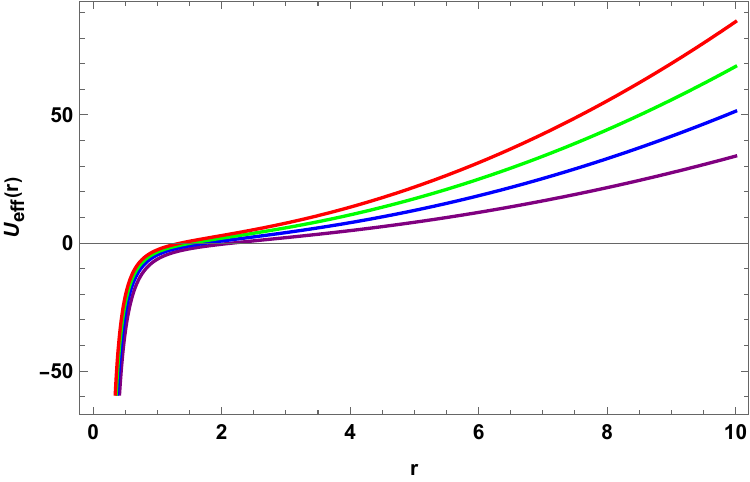}}
\centering{}\caption{The effective potential of time-like geodesics within $\mathcal{RI}$-gravity. Here we set $M=1/2, \Lambda=-0.5, \mathrm{p}_z=0.1$.}\label{fig:4}
\end{figure}
\par\end{center}

In Figures \ref{fig:5} and \ref{fig:6}, we illustrate the effective potential for both null and time-like geodesics in $f(\mathcal{R})$-gravity framework. These figures showcases the variations of $\mathrm{U}_{eff}$ for different values of the coupling constants $\alpha_2,\alpha_3$ and $\alpha_4$. The dotted red line in Figures \ref{fig:5}(a) and \ref{fig:6}(a) represents the cases in general relativity, while the solid lines correspond to various values of the coupling constants for null and time-like geodesics. In Figures \ref{fig:5}(b) and \ref{fig:6}(b), we specify the coupling constants for various color curves as follows: Purple: $\alpha_2=0.1=\alpha_3$, and $\alpha_4=0.0010$; Blue: $\alpha_2=0.2=\alpha_3$, and $\alpha_4=0.0011$; Green: $\alpha_2=0.3=\alpha_3$, and $\alpha_4=0.0012$; and Red: $\alpha_2=0.4=\alpha_3$, and $\alpha_4=0.0014$. These visualizations underscores the impact of these coupling constants on the effective potential of the system within the framework of $f(\mathcal{R})$ gravity theory.

In Figures \ref{fig:3} and \ref{fig:4}, we illustrate the effective potential for both null and time-like geodesics in $\mathcal{RI}$-gravity framework. These figures showcases the variations of $\mathrm{U}_{eff}$ for different values of the coupling constants $\alpha_2,\beta_1,\beta_2$ and $\gamma$. The dotted red line in Figures \ref{fig:3}(a) and \ref{fig:4}(a) represents the cases in general relativity, while the solid lines correspond to various values of the coupling constants for null and time-like geodesics. In Figures \ref{fig:3}(b) and \ref{fig:4}(b), we specify the coupling constants for various color curves as follows: Purple: $\alpha_2=0.001$, $\beta_1=0.1$, $\beta_2=0.001$, and $\gamma=0.001$; Blue: $\alpha_2=0.002$, $\beta_1=0.2$, $\beta_2=0.002$, and $\gamma=0.002$; Green: $\alpha_2=0.003$, $\beta_1=0.3$, $\beta_2=0.003$, and $\gamma=0.003$; and Red: $\alpha_2=0.004$, $\beta_1=0.4$, $\beta_2=0.004$, and $\gamma=0.004$. These visualizations underscores the impact of these coupling constants on the effective potential of the system within the framework of Ricci-Inverse gravity theory.

Now, we will try to solve the above geodesics equations and analyse the result. For null geodesics, $\varepsilon=0$, from Eq. (\ref{C7}), we obtain
\begin{eqnarray}
    \dot{r}=\sqrt{\mathrm{a}+\mathrm{b}\,r},\label{C9}
\end{eqnarray}
where $\mathrm{a}$ and $\mathrm{b}$ are defined as
\begin{eqnarray}
\mathrm{a}=\mathrm{L}^2\,\left(\frac{1}{\beta^2}+\frac{\Lambda_m}{3}\right)-\mathrm{p}^2_{z},\quad\quad \mathrm{b}=\frac{4\,M}{\sqrt{\frac{-\Lambda_m}{3}}}\,\left(1+\frac{3}{\Lambda_m}\,\mathrm{p}^2_{z}\right),\quad \mbox{and}\quad \beta=\frac{\mathrm{L}}{\mathrm{E}}\label{C10}
\end{eqnarray}
is called the impact parameter.

\vspace{0.2cm}
{\bf Case A:} To solve the null geodesics path, let us impose the following constraints on the parameters $\mathrm{a}$ and $\mathrm{b}$ given by
\begin{eqnarray}
    &&\mathrm{a}>0\Rightarrow \frac{\mathrm{L}^2}{\beta^2} >\left(\mathrm{L}^2\,\frac{(-\Lambda_m)}{3}+\mathrm{p}^2_{z}\right),\label{C15}\\
    &&\mathrm{b}>0 \Rightarrow 0 \leq \frac{3\,\mathrm{p}^2_{z}}{(-\Lambda_m)}<1.\label{C16}
\end{eqnarray}

In that case, the solution of Eq. (\ref{C9}) gives us the radial coordinate as follows:
\begin{equation}
    r(\tau)=\frac{1}{\mathrm{b}}\,\left(\frac{\mathrm{b}^2\,(\tau+\mathrm{c}_0)^2}{4}-\mathrm{a}\right)=\frac{\mathrm{b}}{4}\,(\tau+\mathrm{c}_0)^2-\frac{\mathrm{a}}{\mathrm{b}},\label{C11}
\end{equation}
where $\mathrm{c}_0$ is a constant of integration. Now, as the affine parameter approaches zero, the radial coordinate $r(\tau) \geq 0$. Thus, at $\tau=0$ we have $r(\tau=0) \geq 0$ which implies 
\begin{equation}
    \mathrm{c}_0 \geq \frac{2\,\sqrt{\mathrm{a}}}{\mathrm{b}} \quad \mbox{since}\quad \mathrm{a}>0.
\end{equation}

Substituting $r(\tau)$ from Eq. (\ref{C11}) into the Eq. (\ref{C4}) and after integration, we obtain
\begin{equation}
    \varphi(\tau)=\frac{\mathrm{b}^2\,\mathrm{L}}{\mathrm{a}}\,\left[-\frac{(\tau+\mathrm{c}_0)}{2\,\left(\frac{\mathrm{b}^2\,(\tau+\mathrm{c}_0)^2}{4}-\mathrm{a}\right)}+\frac{\tanh^{-1}\left(\frac{\mathrm{b}\,\tau}{2\,\sqrt{\mathrm{a}}}+1\right)}{\mathrm{b}\,\sqrt{\mathrm{a}}}  \right]+\mathrm{c}_1,\label{C12}
\end{equation}
where $\mathrm{c}_1$ is an arbitrary constant.

Similarly, substituting $r(\tau)$ from Eq. (\ref{C11}) into the Eq. (\ref{C5}) and after integration, we obtain
\begin{equation}
    z(\tau)=\frac{3\,\mathrm{b}^2\,\mathrm{p}_z}{\mathrm{a}\,(-\Lambda_m)}\,\left[-\frac{(\tau+\mathrm{c}_0)}{2\,\left(\frac{\mathrm{b}^2\,(\tau+\mathrm{c}_0)^2}{4}-\mathrm{a}\right)}+\frac{\tanh^{-1}\left(\frac{\mathrm{b}\,\tau}{2\,\sqrt{\mathrm{a}}}+1\right)}{\mathrm{b}\,\sqrt{\mathrm{a}}}  \right]+\mathrm{c}_2,\label{C13}
\end{equation}
where $\mathrm{c}_2$ is an arbitrary constant. 

\vspace{0.2cm}
{\bf Case B:} To solve the above null geodesics path, let us impose the following constraints on the parameters $\mathrm{a}$ and $\mathrm{b}$ given by
\begin{eqnarray}
    &&\mathrm{a}=0\Rightarrow \frac{\mathrm{L}^2}{\beta^2} =\left(\mathrm{L}^2\,\frac{(-\Lambda_m)}{3}+\mathrm{p}^2_{z}\right),\label{D1}\\
    &&\mathrm{b}>0 \Rightarrow 0 \leq \frac{3\,\mathrm{p}^2_{z}}{(-\Lambda_m)}<1.\label{D11}
\end{eqnarray}

In that case, the solution of Eq. (\ref{C9}) gives us the radial coordinate as follows:
\begin{equation}
    r(\tau)=\frac{\mathrm{b}}{4}\,(\tau+\mathrm{k}_0)^2,\label{D2}
\end{equation}
where $\mathrm{k}_0$ is a constant of integration. Employing the similar argument on the radial coordinate $r(\tau=0)> 0$ as done earlier, we have $\mathrm{k}_0 > 0$.

Therefore, the geodesics paths for $\varphi(\tau)$ and $z(\tau)$ are given by
\begin{eqnarray}
    &&\varphi(\tau)=-\frac{16\,\mathrm{L}}{3\,\mathrm{b}^2}\,\frac{1}{(\tau+\mathrm{k}_0)^3}+\mathrm{k}_1,\label{D3}\\
    &&z(\tau)=-\frac{16\,\mathrm{p}_z}{\mathrm{b}^2\,(-\Lambda_m)}\,\frac{1}{(\tau+\mathrm{k}_0)^3}+\mathrm{k}_2.\label{D4}
\end{eqnarray}

\vspace{0.2cm}
{\bf Case C:} Let us impose the following constraints on the parameters $\mathrm{a}$ and $\mathrm{b}$ given by
\begin{eqnarray}
    &&\mathrm{a}<0\Rightarrow \frac{\mathrm{L}^2}{\beta^2} <\left(\mathrm{L}^2\,\frac{(-\Lambda_m)}{3}+\mathrm{p}^2_{z}\right),\quad \mathrm{a}=-\bar{\mathrm{a}},\quad \bar{\mathrm{a}}=\mathrm{L}^2\,\left(-\frac{1}{\beta^2}+\frac{(-\Lambda_m)}{3}\right)+\mathrm{p}^2_{z}>0,\label{EE}\\
    &&\mathrm{b}>0 \Rightarrow 0 \leq \frac{3\,\mathrm{p}^2_{z}}{(-\Lambda_m)}<1.\label{E1}
\end{eqnarray}

In that case, the solution of Eq. (\ref{C9}) gives us the radial coordinate as follows:
\begin{equation}
    r(\tau)=\frac{\mathrm{b}}{4}\,(\tau+\mathrm{d}_0)^2+\frac{\mathrm{\bar{a}}}{\mathrm{b}}=\frac{\mathrm{b}\,\tau^2}{4}+\frac{\mathrm{\bar{a}}}{\mathrm{b}}=\frac{1}{\mathrm{b}}\,\left(\frac{\mathrm{b}^2\,\tau^2}{4}+\mathrm{\bar{a}}\right),\label{E2}
\end{equation}
where $\mathrm{d}_0$ is a constant of integration, which we set equal to zero. Now, as the affine parameter approaches zero, the radial coordinate satisfies $r(\tau \geq 0) \geq \kappa$, where $\kappa=\frac{\mathrm{\bar{a}}}{\mathrm{b}}$. 

Substituting $r(\tau)$ from Eq. (\ref{E2}) into Eq. (\ref{C4}) and integrating, we obtain:
\begin{equation}
    \varphi(\tau)=\frac{\mathrm{b}^2\,\mathrm{L}}{\mathrm{\bar{a}}}\,\left[-\frac{\tau}{2\,\left(\frac{\mathrm{b}^2\,\tau^2}{4}+\mathrm{\bar{a}}\right)}+\frac{\tanh^{-1}\left(\frac{\mathrm{b}\,\tau}{2\,\sqrt{\mathrm{\bar{a}}}}\right)}{\mathrm{b}\,\sqrt{\mathrm{\bar{a}}}}  \right]+\mathrm{d}_1,\label{E4}
\end{equation}
where $\mathrm{d}_1$ is an arbitrary constant.

Similarly, substituting $r(\tau)$ from Eq. (\ref{E2}) into the Eq. (\ref{C5}) and after integration, we obtain
\begin{equation}
    z(\tau)=\frac{3\,\mathrm{b}^2\,\mathrm{p}_z}{\mathrm{\bar{a}}\,(-\Lambda_m)}\,\left[-\frac{\tau}{2\,\left(\frac{\mathrm{b}^2\,\tau^2}{4}+\mathrm{\bar{a}}\right)}+\frac{\tanh^{-1}\left(\frac{\mathrm{b}\,\tau}{2\,\sqrt{\mathrm{\bar{a}}}}\right)}{\mathrm{b}\,\sqrt{\mathrm{\bar{a}}}}  \right]+\mathrm{d}_2,\label{E5}
\end{equation}
where $\mathrm{d}_2$ is an arbitrary constant. 

\begin{table}[ht!]
    \centering
    \begin{tabular}{c|c|c}
    \hline\hline
    Case &  Range of $r$ & $r(\tau)$, $\varphi(\tau)$, $z(\tau)$ \\[1ex]
    \hline\hline
    $\epsilon=0$, $\frac{\mathrm{L}^2}{\beta^2}>\mathrm{L}^2\,\frac{(-\Lambda_m)}{3}+p^2_{z}$ & $0 \leq r < \infty$ & $\begin{array}{c}
         r(\tau)=\frac{\mathrm{b}}{4}\,(\tau+c_0)^2-\frac{\mathrm{a}}{\mathrm{b}}
         \hfill\\  [2ex]
        \varphi(\tau)=\frac{\mathrm{b}^2\,\mathrm{L}}{\mathrm{a}}\,\left[-\frac{(\tau+\mathrm{c}_0)}{2\,\left(\frac{\mathrm{b}^2\,(\tau+\mathrm{c}_0)^2}{4}-\mathrm{a}\right)}+\frac{\tanh^{-1}\left(\frac{\mathrm{b}\,\tau}{2\,\sqrt{\mathrm{a}}}+1\right)}{\mathrm{b}\,\sqrt{\mathrm{a}}}\right]
        \hfill\\  [2ex]
        z(\tau)=\frac{3\,\mathrm{b}^2\,\mathrm{p}_z}{\mathrm{a}\,(-\Lambda_m)}\,\left[-\frac{(\tau+\mathrm{c}_0)}{2\,\left(\frac{\mathrm{b}^2\,(\tau+\mathrm{c}_0)^2}{4}-\mathrm{a}\right)}+\frac{\tanh^{-1}\left(\frac{\mathrm{b}\,\tau}{2\,\sqrt{\mathrm{a}}}+1\right)}{\mathrm{b}\,\sqrt{\mathrm{a}}}  \right]
    \end{array}$\\[6ex]
    \hline 
    $\epsilon=0$, $\frac{\mathrm{L}^2}{\beta^2} =\left(\mathrm{L}^2\,\frac{(-\Lambda_m)}{3}+\mathrm{p}^2_{z}\right)$ & $0 < r < \infty$  & $\begin{array}{c}
         r(\tau)=\frac{\mathrm{b}}{4}\,(\tau+\mathrm{k}_0)^2
         \hfill\\ [2ex]
         \varphi(\tau)=-\frac{16\,\mathrm{L}}{3\,\mathrm{b}^2}\,\frac{1}{(\tau+\mathrm{k}_0)^3}
         \hfill\\ [2ex]
         z(\tau)=-\frac{16\,\mathrm{p}_z}{\mathrm{b}^2\,(-\Lambda_m)}\,\frac{1}{(\tau+\mathrm{k}_0)^3}
    \end{array}$\\[6ex]
    \hline 
    $\epsilon=0$, $\frac{\mathrm{L}^2}{\beta^2} <\left(\mathrm{L}^2\,\frac{(-\Lambda_m)}{3}+\mathrm{p}^2_{z}\right)$ & $\kappa \leq r < \infty$ & $\begin{array}{c}
         r(\tau)=\frac{1}{\mathrm{b}}\,\left(\frac{\mathrm{b}^2\,\tau^2}{4}+\mathrm{\bar{a}}\right)
         \hfill\\  [2ex]
        \varphi(\tau)=\frac{\mathrm{b}^2\,\mathrm{L}}{\mathrm{\bar{a}}}\,\left[-\frac{\tau}{2\,\left(\frac{\mathrm{b}^2\,\tau^2}{4}+\mathrm{\bar{a}}\right)}+\frac{\tanh^{-1}\left(\frac{\mathrm{b}\,\tau}{2\,\sqrt{\mathrm{\bar{a}}}}\right)}{\mathrm{b}\,\sqrt{\mathrm{\bar{a}}}}  \right]
        \hfill\\  [2ex]
        z(\tau)=\frac{3\,\mathrm{b}^2\,\mathrm{p}_z}{\mathrm{\bar{a}}\,(-\Lambda_m)}\,\left[-\frac{\tau}{2\,\left(\frac{\mathrm{b}^2\,\tau^2}{4}+\mathrm{\bar{a}}\right)}+\frac{\tanh^{-1}\left(\frac{\mathrm{b}\,\tau}{2\,\sqrt{\mathrm{\bar{a}}}}\right)}{\mathrm{b}\,\sqrt{\mathrm{\bar{a}}}}  \right]
    \end{array}$\\[6ex]
    \hline \hline
\end{tabular}
\caption{Null geodesic equations for LBH space-time within the modified gravity theories.}
\label{tab:1}
\end{table}

Table \ref{tab:1} outlines the null geodesic equations for a black hole space-time (LBH) within modified gravity theories, organized into three distinct cases based on the relationship between the parameters $\frac{\mathrm{L}^2}{\beta^2}$, $\mathrm{L}^2\,\frac{(-\Lambda_m)}{3}$, and $\mathrm{p}^2_{z}$. Each case specifies the range of the radial coordinate $r$ and provides explicit solutions for $r(\tau)$, $\varphi(\tau)$, and $z(\tau)$, which describe the trajectories of light-like particles in this space-time. In the first case, where $\frac{\mathrm{L}^2}{\beta^2} > \mathrm{L}^2\,\frac{(-\Lambda_m)}{3} + \mathrm{p}^2_{z}$, the radial coordinate $r$ extends from 0 to infinity, and the solutions for $\varphi(\tau)$ and $z(\tau)$ involve inverse hyperbolic tangent functions, reflecting a specific dependence on the parameter $\tau$. The second case, where $\frac{\mathrm{L}^2}{\beta^2} = \mathrm{L}^2\,\frac{(-\Lambda_m)}{3} + \mathrm{p}^2_{z}$, also spans $0 < r < \infty$, but the solutions simplify significantly, with $\varphi(\tau)$ and $z(\tau)$ exhibiting inverse cubic dependence on $\tau$. In the third case, where $\frac{\mathrm{L}^2}{\beta^2} < \mathrm{L}^2\,\frac{(-\Lambda_m)}{3} + \mathrm{p}^2_{z}$, the radial coordinate $r$ is constrained to $\kappa \leq r < \infty$, and the solutions again incorporate inverse hyperbolic tangent functions, albeit with a different parametric structure. These results demonstrate how the interplay between angular momentum, cosmological constant, and linear momentum in modified gravity theories shapes the behavior of null geodesics, offering insights into the underlying geometry and dynamics of the LBH space-time.

From the above null geodesics analysis, it is evident that the null geodesics paths are finite and smooth throughout the interval $\tau \in (-\infty, \infty)$. We summarize the null geodesics in Table \ref{tab:1}. As mentioned earlier, in the $f(\mathcal{R})$-gravity framework, $\Lambda_m$ is defined by $\Lambda^{f(\mathcal{R})}_m$ as presented in Eq. (\ref{CC7}). Similarly, in Ricci-Inverse gravity, the parameter $\Lambda_m$ transforms to $\Lambda^{\mathcal{RI}}_m$ as outlined in Eq. (\ref{B14}). Thus, the trajectories of photon light rays are influenced by the constants $\alpha_2$, $\alpha_3$, and $\alpha_4$ coupled with higher-order curvature $\mathcal{R}$ in $f(\mathcal{R})$ gravity theory. Similarly, the photon trajectories are influenced by the coupling constants $\alpha_2$, $\beta_1$, $\beta_2$, and $\gamma$ in Ricci-Inverse gravity.

\subsection{Force on the massless photon particle}

In this part, we determine the force acting on the massless photon particle within the framework of modified gravity theories. Specifically, we focus on $f(\mathcal{R})$-gravity and Ricci-Inverse gravity, demonstrating how the coupling constants influence the force on the massless photon particle.

The force on the massless photon particle is defined by
\begin{equation}
    F=-\frac{1}{2}\,\frac{dU_\text{eff}}{dr}.\label{HH1}
\end{equation}
Using the expression (\ref{C888}), we find this force in general relativity case as,
\begin{equation}
    F=\frac{\Lambda}{3}\,r-\frac{2\,M}{r^2\,\sqrt{-\frac{\Lambda}{3}}}.\label{HH2}
\end{equation}

\begin{figure}[ht!]
    \centering
    \includegraphics[width=0.47\linewidth]{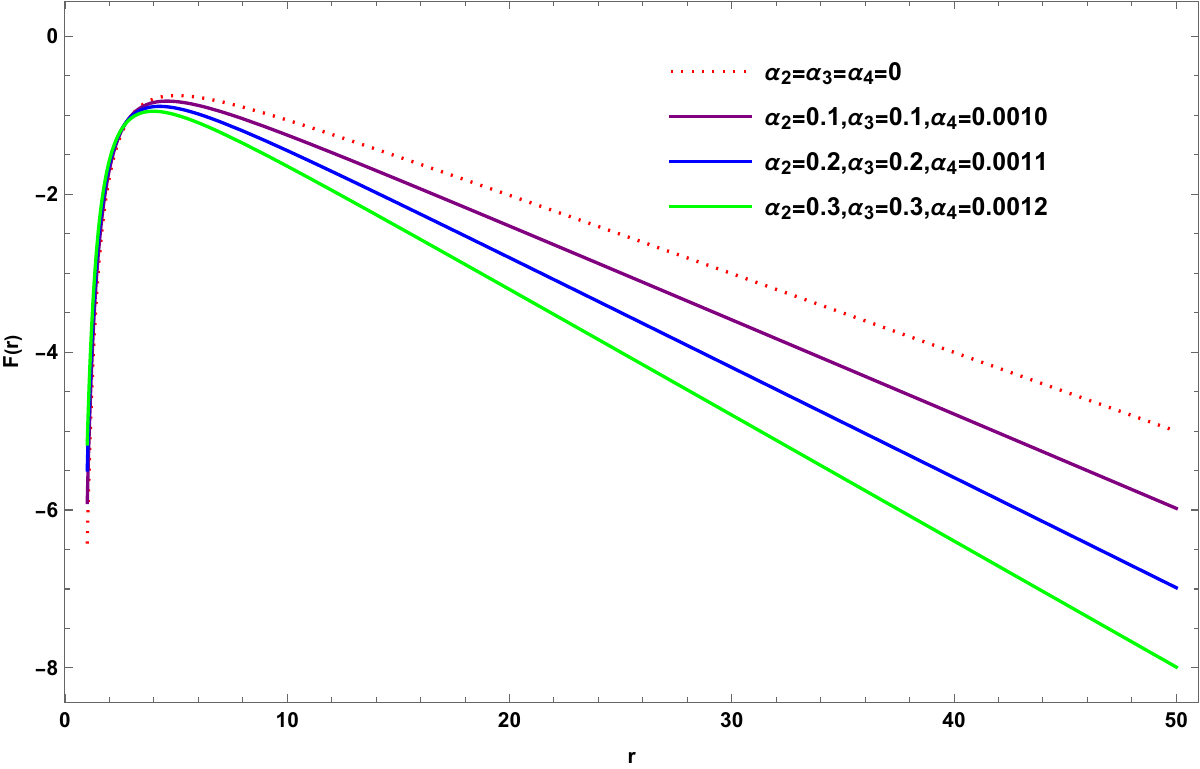}\quad
    \includegraphics[width=0.47\linewidth]{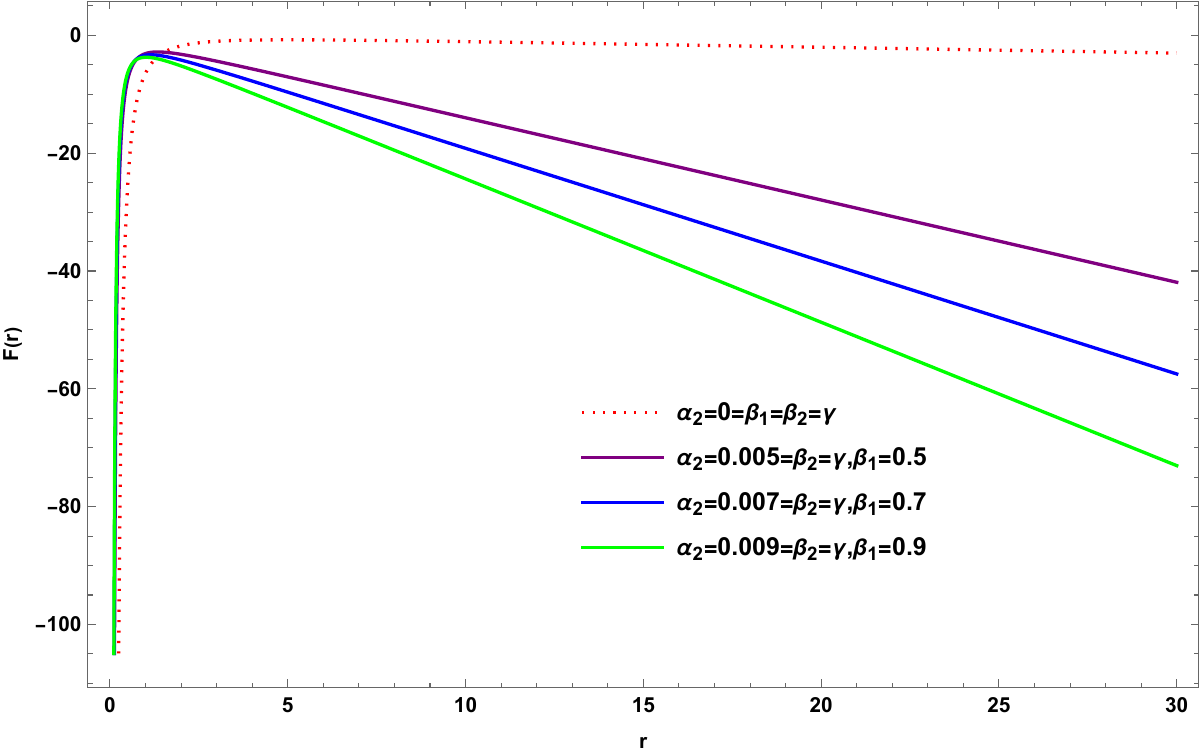}
    \caption{The force on massless photon particles in $f(\mathcal{R})$ gravity (left panel) and Ricci-Inverse gravity (right panel). Here $M=1$ and the usual cosmological constant $\Lambda=-0.3$.}
    \label{fig:7}
\end{figure}

Therefore, in the context of $f(\mathcal{R})$ gravity, this force will be
\begin{equation}
    F_{f(\mathcal{R})}=\left(\frac{\Lambda-16\Lambda^3\alpha_2-128\Lambda^4\alpha_3-768\alpha_4\Lambda^5}{3}\,r-\frac{2\,M}{r^2\,\sqrt{-\frac{(\Lambda-16\Lambda^3\alpha_2-128\Lambda^4\alpha_3-768\alpha_4\Lambda^5)}{3}}}  \right).\label{HH3}
\end{equation}
While in the framework of Ricci-Inverse gravity, we find
\begin{equation}
    F_{\mathcal{RI}}=\left(\frac{\Lambda+16\Lambda^3\alpha_2+\frac{3\beta_1}{\Lambda}+\frac{16\beta_2}{\Lambda^2}+\frac{4\gamma}{\Lambda^2}}{3}\,r-\frac{2\,M}{r^2\,\sqrt{-\frac{\left(\Lambda+16\Lambda^3\alpha_2+\frac{3\beta_1}{\Lambda}+\frac{16\beta_2}{\Lambda^2}+\frac{4\gamma}{\Lambda^2}\right)}{3}}}  \right).\label{HH4}
\end{equation}

From the above expressions (\ref{HH3}) and (\ref{HH4}), it is evident that the force on massless photon light is altered in the modified gravity theories compared to the result in the general relativity case. This force is influenced by the constants $\alpha_2$, $\alpha_3$, and $\alpha_4$ coupled with the higher-order scalar curvature $\mathcal{R}$ in $f(\mathcal{R})$ gravity theory. Similarly, the force is influenced by the coupling constants $\alpha_2$, $\beta_1$, $\beta_2$, and $\gamma$ in Ricci-Inverse gravity.

In Figure (\ref{fig:7}), we illustrate the force on the massless photon particle in modified gravity theories, specifically in the $f(\mathcal{R})$-gravity (left panel) and Ricci-Inverse gravity (right panel) frameworks. The red dotted lines correspond to the general relativity case, while the colored lines represent the modified gravity frameworks. These visualizations underscore the impact of the coupling constants on the force experienced by massless photon particles within the framework of modified gravity theories, in comparison to the results in the general relativity case.

\section{Conclusions}\label{sec:5}

This study provided a detailed exploration of the Lemos cylindrical black hole (LBH) spacetime within the framework of modified gravity theories, specifically focusing on $f(\mathcal{R})$-gravity, where we considered higher-order curvature terms of the form $f(\mathcal{R})=(\mathcal{R}+\alpha_1\,\mathcal{R}^2+\alpha_2\,\mathcal{R}^3+\alpha_3\,\mathcal{R}^4+\alpha_4\,\mathcal{R}^5)$. This extension allowed for an exploration of non-linear modifications to the Ricci scalar, which were expected to become relevant in regimes of strong gravitational fields, such as near black hole horizons. We derived the modified field equations by considering the Lemos black hole and solved them with vacuum as the stress-energy tensor. In fact, we observed that the coupling constants $\alpha_i$ influenced or shifted the usual constants, thereby altering the geometric structure.

Furthermore, we extended our analysis to another modified gravity framework known as Ricci-Inverse ($\mathcal{RI}$) gravity. By analyzing three distinct classes of models in Ricci-Inverse gravity, we investigated how the inclusion of an anti-curvature tensor and its scalar altered the geometric structure of black holes. In fact, we demonstrated that the considered cylindrical black hole represented valid solutions across all Class-{\bf I} to Class-{\bf III} models within Ricci-Inverse gravity. Additionally, we solved the modified field equations by adopting vacuum as the energy-momentum tensor and obtained an expression for the effective cosmological constant in terms of the usual cosmological constant $\Lambda$.

Finally, we studied the geodesic motion of test particles and light trajectories around the Lemos black hole (LBH) spacetime within the $f(\mathcal{R})$ and Ricci-Inverse gravity frameworks, which provided further insights into how modified gravity affects the dynamics of massive and massless particles. Our analysis showed that test particles moving in the vicinity of this black hole experienced altered gravitational forces depending on the specific modified gravity model employed. This suggested that the modified gravity framework introduced additional forces that could influence particle orbits, the stability of circular orbits, and even the nature of event horizons. These results are not only relevant for understanding black hole physics but also for potential observational tests of modified gravity, such as tracking the motion of stars or gas clouds near supermassive black holes.

This paper provided a comprehensive study of the cylindrical Lemos black hole spacetime within the context of $f(\mathcal{R})$ and Ricci-Inverse gravity frameworks, shedding light on how modifications to the gravitational field equations influenced the geometric structure and particle dynamics. Our results highlighted the impact of different coupling constants $\alpha_i, \beta_i$, and $\gamma$ on the structure and dynamics of black holes. The findings of this study underscored the importance of exploring alternative gravity theories as a means to deepen our understanding of black hole physics, cosmology, and potentially the quantum nature of gravity itself. In our next work, we plan to focus on the thermodynamics and quasinormal modes of black hole spacetime constructed within the modified gravity theories discussed here.

\section*{Acknowledgments}

F.A. acknowledges the Inter University Centre for Astronomy and Astrophysics (IUCAA), Pune, India for granting visiting associateship.

\section*{Data Availability Statement}

This manuscript has no associated data. [Authors’ comment: All data generated or analyzed during this study are included in this published article.]

\section*{Code Availability Statement}

This manuscript has no associated code/software. [Authors’ comment: Code/Software sharing not applicable as no code/software was generated during in this current study.]

\section*{Conflict of Interests}

Author declare(s) no conflict of interest.

\end{document}